\documentclass[11pt,a4paper]{report}
\pdfoutput=1
\usepackage{caption, subcaption, amsmath, url, pifont, graphicx, cite, booktabs, multirow, rotating, array, mathtools, bm, threeparttable}
\DeclarePairedDelimiter\ceil{\lceil}{\rceil}
\DeclarePairedDelimiter\floor{\lfloor}{\rfloor}

\makeatletter
\def\endthebibliography{%
	\def\@noitemerr{\@latex@warning{Empty `thebibliography' environment}}%
	\endlist
}
\makeatother

\title{\huge Real-time Tone Mapping:\\A State of the Art Report}

\author{Yafei~Ou $^1$ $^3$ $^6$ \and 
	Prasoon~Ambalathankandy \thanks{Equally Contributed} $^3$ $^6$ \and 
	Masayuki~Ikebe \thanks{Corresponding author} \thanks{Y. Ou, P.Ambalathankandy and M.Ikebe are with Research Center For Integrated Quantum Electronics, Hokkaido University, Sapporo, Japan e-mail: \texttt{ikebe@ist.hokudai.ac.jp}}\and 
	Shinya~Takamaeda \thanks{S.Takamaeda is with Department of Computer Science, Graduate School of Information Science and Technology, University of Tokyo, Tokyo, Japan} \and 
	Masato~Motomura \thanks{M.Motomura is with Institute of Innovative Research, Tokyo Institute of Technology, Tokyo, Japan} \and 
	Tetsuya~Asai \thanks{Y. Ou, P.Ambalathankandy and T.Asai are with Graduate School of Information Science and Technology, Hokkaido University, Sapporo, Japan}}
	
\begin{document}


\bibliographystyle{IEEEtranTIE}

\maketitle

\begin{abstract}
The rising demand for high quality display has ensued active research in high dynamic range (HDR) imaging, which has the potential to replace the standard dynamic range imaging. This is due to HDR's features like accurate reproducibility of a scene with its entire spectrum of visible lighting and color depth. But this capability comes with expensive capture, display, storage and distribution resource requirements. Also, display of HDR images/video content on an ordinary display device with limited dynamic range requires some form of adaptation. Many adaptation algorithms, widely known as tone mapping operators, have been studied and proposed in the last few decades. In this state of the art report, we present a comprehensive survey of $50+$ tone mapping algorithms that have been implemented on hardware for acceleration and real-time performance. These algorithms have been adapted or redesigned to make them hardware-friendly. This effort leads to various design challenges that are encountered during the hardware development. Any real-time application poses strict timing constraints which requires time exact processing of the algorithm. Also, most of the embedded systems would have limited system resources in terms of battery, computational power and memory resources. These design challenges require novel solutions, and in this report we focus on these issues.    

In this we survey will discuss those tonemap algorithms which have been implemented on GPU \protect \cite{goodnight2005interactive, krawczyk2005perceptual, roch2007interactive, zhao2008real, tiant2012gpu, akil2012real, urena2013real, eilertsen2015real, khan2017tone, tsai2019real}, FPGA \cite{hassan2007fpga, marsi2007video, iakovidou2008fpga, vakili2011customized, lapray2011smart, lapray2012hdr, kiser2012real, urena2012real, mann2012realtime, lapray20131, ofili2013hardware, vytla2013real, canada2013embedded, popovic2014performance, shiau2014low, li2015low, lapray2016hdr, ambalathankandy2019fpga, shahnovich2016hardware, liu2016study, popovic2016multi, li2016novel, nosko2017true, zemvcik2017real, popadic2017method, nosko2018color, yang2018local, yang2019mantissa, liu2019high, ambalathankandy2019adaptive, park2019low}, and ASIC \cite{chiu2011real, punchihewa2011review, sicard2013cmos, vargas2014151, gouveia2014reconfigurable, mughal2014threshold, mughal2015fixed, fernandez2015single, shi2016tone, shi2016analog, chen2016analog, guicquero2016algorithm} in terms of their hardware specifications and performance. Output image quality is an important metric for tonemap algorithms. From our literature survey we found that, various objective quality metrics have been used to demonstrate the functionality of adapting the algorithm on hardware platform. We have compiled and studied all the metrics used in this survey \cite{shannon1948mathematical, beghdadi1989contrast, cameron1997r, wang2002universal, chen2003minimum, wang2004image, smith2006beyond, agaian2007transform, panetta2008human, mukherjee2008enhancement, watkins2010statistics, celik2011contextual, yeganeh2012objective, tsai2012fast}. Finally, in this report we demonstrate the link between hardware cost and image quality thereby illustrating the underlying trade-off. This report concludes with a discussion on the general future research directions based-on various hardware design/implementation bottlenecks which will be useful for the research community.

\end{abstract}
	

\chapter{Introduction}

\begin{figure*}
\includegraphics [width=\textwidth]{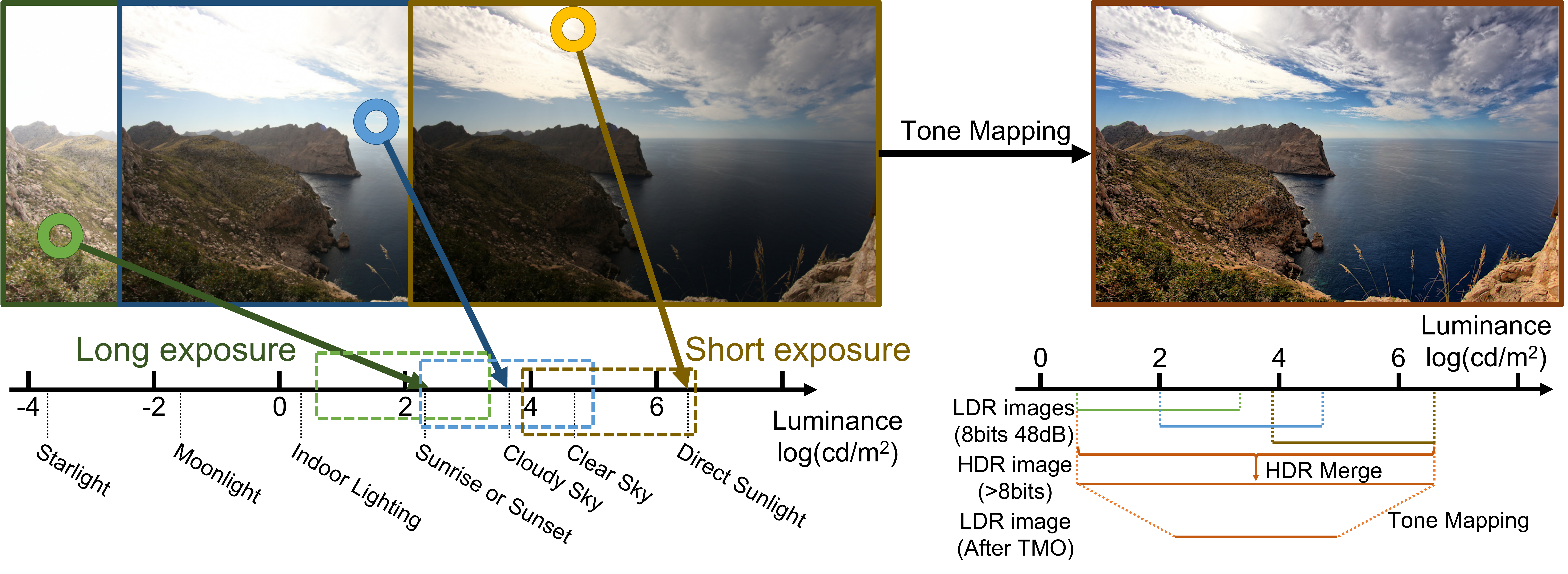}
\caption{This figure illustrates how multiple exposure images are required to capture the wide ambient luminance levels that exists in our natural lighting environments. The HDR images produced from these multi-exposure images have higher bit-depths and a TMO is required to faithfully display it on a common display device (images from \cite{easyHDR}).}
\label{fig:teaser1}
\end{figure*}

Superior display quality is a dominant feature that has been driving the consumer electronics industry. Unlike the past, the growing demand for definitive viewing experience is not only limited to entertainment, gaming, and media industry but has been increasingly sought for applications in security and surveillance, automotive, medical, AR-VR, drones and robotics imaging systems. High Dynamic Range (HDR) imaging has come to become a compelling aspect of the new 4K/8K Ultra-high-definition (UHD) format \cite{Monitors}. Luminance and contrast, which are very important to the human eye, and our modern display systems have problems when dealing with visuals that may have details simultaneously both in sun, and in shadows. HDR imaging looks to solve these problems. This technique accounts for more realistically contrasted visuals by bringing out colors and detail in low-light areas so that visuals in shadow are not compressed, while not saturating the highlights. In other words, HDR can make the dark visuals deeper and the lights brighter, with more color shades with optimized contrast ratio of the display. However, this increase in amount of detail and extended color space comes at the price of higher data-width, thereby it requires more hardware/software resources to create, distribute and display HDR content.

Dynamic range of a digital images is defined as the ratio between the darkest and the brightest points captured from a scene \cite{reinhard2010high}. It can be expressed in orders of magnitude (powers of ten), in stops (powers of two) or in decibels (db). From Fig\ref{fig:teaser1}, we can notice that our eyes can see objects both in a dark night and in a sunny day, although the luminance level of a scene in sunlight is about $10^{6}cd/m^{2}$ and one with starlight is about $10^{-3}cd/m^{2}$ \cite{banterle2017advanced}. This means that our human visual system (HVS) is capable of adapting to wide lighting variations within range of nearly 10 orders of magnitude. From \cite{reinhard2010high}, it is learnt that HVS can easily adapt  up to 5 orders of magnitude within the same scene. HDR imaging aims to increase the dynamic range recorded in a digital image from a given scene. Pixels in an HDR image are proportional to the radiance of the scene, dark and bright areas can be recorded within the same image. But one of the main limitations of our digital cameras is their inability in capturing and storing the HDR of the natural scenes. Which visually implies under and over exposure in bright and dark regions. Ordinary digital cameras produce images in a range lower than $1:1000 cd/m^{2}$ \cite{jacobs2008automatic}. There by most of the digital images are still Low Dynamic Range (LDR), with a data-width of 24 bits per pixel (in RGB format 8-bits per channel). Which translates to approximately 2 orders of magnitude while HDR images may have 80 orders dynamic range represented in floating point formats \cite{reinhard2010high}. Common displays or printers represents only 8 bits per color channel (approximately $1:300 cd/m^{2}$) \cite{seetzen2004high}, therefore HDR images need to be adapted (i.e., tone mapped) to 8 bits per color channel to display or print them in LDR devices. Figure \ref{fig:teaser1} shows a tone mapped example of HDR image, notice that both bright and dark regions of the image are properly displayed.

There is a wide gap between the range of light that we can see \& perceive and what a common digital camera can capture and display on a monitor respectively. Ordinary digital camera's image sensors have limited capability for recording the entire illumination range in a natural scene. Also it is a challenge to determine good exposure values, especially in diverse scenes which have good  large dark and bright areas. For example, taking a picture on a sunny day we have to chose whether to appropriately expose the bright sky and also account for the details in the shadow of the mountain slopes like in Fig \ref{fig:teaser1}. Modern digital cameras come with in-built auto-exposure algorithm to automatically set the ISO value, aperture and shutter speed corresponding to the given scene. However, a scene saturation (over-exposure) can occur if the scene is very brightly lit (direct sunlight) and sensor records the maximum allowed value, therefore details in bright areas are clamped to the maximum allowed value which is white. On the other hand, if the scene is poorly lighted and the light energy reaching the sensor is very low it results in under-exposed image. As stated earlier, we have to adapt the image so that we can match the dynamic range of HDR scene with the display device's dynamic range, this process is widely known as tone mapping. Depending upon the dynamic range of the captured image tone mapping function can expand or compresses it in order to enhance the display quality \cite{meylan2006tone}. The purpose of applying tone mapping on an image can be different and depends on the particular application. In some cases it may be to improve the aesthetics of the image, while for another application it might be to emphasize as many details as possible, or could be to maximize the image contrast \cite{hoefflinger2007high}. However, the ultimate goal of tone mapping is to match the perception of tone mapped image with the real world perception \cite{reinhard2010high}. A tone mapping operator (TMO) $f$ can be defined as a transformation function $f(I)$:
\begin{equation}
 f(I):R^{w\times h \times c} \rightarrow D^{w\times h \times c}
\label{Eq:tm1}
\end{equation}
Here, the tonemap function $f$ maps real world luminance $R$ to display luminance $D$ \cite{banterle2017advanced}, and  $I$ is an image with dimension $w \times h$ and $c$ is number of color bands which is $3$ for RGB image. Tone mapping has been an active area of research for the last two decades, resulting in the design and development of many hundreds of different tone mapping algorithms which can be broadly grouped in to global, local, frequency and segmentation operators
\cite{banterle2017advanced}.

\begin{figure*}
\includegraphics [width=\textwidth]{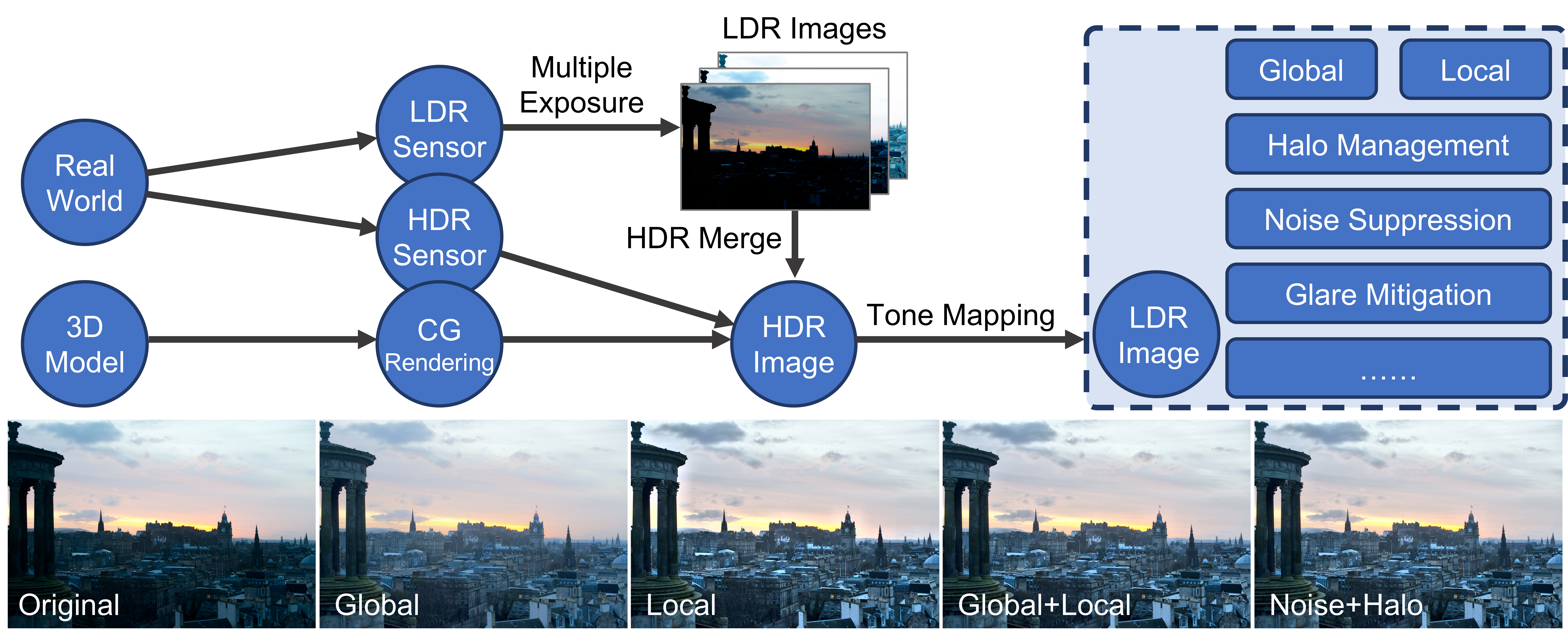}

\caption{HDR imaging and tone mapping: Global tone mapping functions are good for capturing overall preview of the input image. Local tone mapping function by considering pixel neighborhood information for each input pixel, it can emphasize more local details. Additional filters are used to improve the subjective quality of tonemapped images (original image from \cite{Edinburgh}).}
\label{fig:teaser2}
\end{figure*}

Global TMOs, apply the same function to all pixels in the image. These mapping function treats every pixel of the image independently. These operators are computationally efficient, easy to implement and can be executed in real time. A local tone mapping function compresses a pixel value according to its luminance values and their neighboring pixels luminance values. Hence, for each pixel the computation is adjusted according to an average over a local neighborhood of pixels \cite{reinhard2010high}. Frequency domain-based operators, like the local TMO preserve edges and local contrast by computing in the frequency domain instead of spatial \cite{banterle2017advanced, kuang2007icam06}. Segmentation operators divides input image into different uniform regions, and a global mapping operator is applied on each of these regions and finally these are merged to produce output image \cite{lischinski2006interactive, mertens2007exposure}. HDR image pixels can have large disparity in intensities in small neighborhood thereby exhibiting artifacts in tone mapped images (particularly in local tonemap), like in Fig. \ref{fig:teaser2}. Therefore, various filters are required to suppress these artifacts and improve aesthetics of output images depending upon the targeted application. 

In literature, there have been many state of the art survey reports like \cite{matkovic1997survey, devlin2002review, drago2002perceptual, vcadik2008evaluation, rao2012survey, eilertsen2013survey, eilertsen2017comparative}. These surveys mainly covered the software algorithms. The choice of a TMO operator is usually application driven, and in this report we study algorithms that have been optimized for hardware platforms like Application Specific Integrated Circuits (ASIC), Field Programmable Gate Arrays (FPGA) and dedicate Graphic Processing Unit (GPU). Strong demand for real-time embedded vision-based applications is on the rise in various domains like advanced automotive systems, medical imaging, robotics and Unmanned Aerial Vehicles (UAVs). The main building blocks for such vision-based systems are the image sensors, image processing algorithms and display monitors. For real-time applications with time constraints a hardware acceleration is necessary \cite{kalb2016tulipp}. Also, embedded applications are energy and resource constrained there by simply porting software algorithms on a hardware platform may result in poor performance or even system failure. Therefore, image processing algorithms have to be optimized for hardware porting \cite{bailey2011design}. This redesign effort, which can exploit the hardware platform for optimal performance has produced many novel hardware tone mapping algorithms and architectures. In this survey, we report such hardware tone mapping algorithms and to the best of our knowledge there has been no such earlier survey. Following are our main contributions:
\begin{enumerate}
\item A comprehensive introduction to hardware TMOs and imaging pipeline.
\item Detailed survey of TMOs that have been implemented on an ASIC/FPGA and GPU platform.
\item Comparison of TMOs based on their hardware specification and performance.
\item Image quality assessment for hardware TMOs.
\item Demonstrate the link between hardware cost and output image quality.
\item Discussion on future perspectives for implementing machine-learning based TMOs on hardware.
\end{enumerate}

The rest of paper is organized as follows: Section II covers HDR imaging. Section III presents a detailed survey of all hardware TMO algorithms. In section IV we will present the design issues and implementation TMO hardware techniques. We will also present figures of merit for these implementations, thus throwing light on design trade-offs. Image quality assessment is a very important performance metric, and in section V we list all the different metrics that are used in the current literature. While drawing our conclusions in section VI, we also present an account on the future research directions.
\chapter{High Dynamic Range Imaging}\label{HDRI}

\begin{figure}[!b]
\centerline{\includegraphics[width=8cm, trim=0 70 0 10, clip]{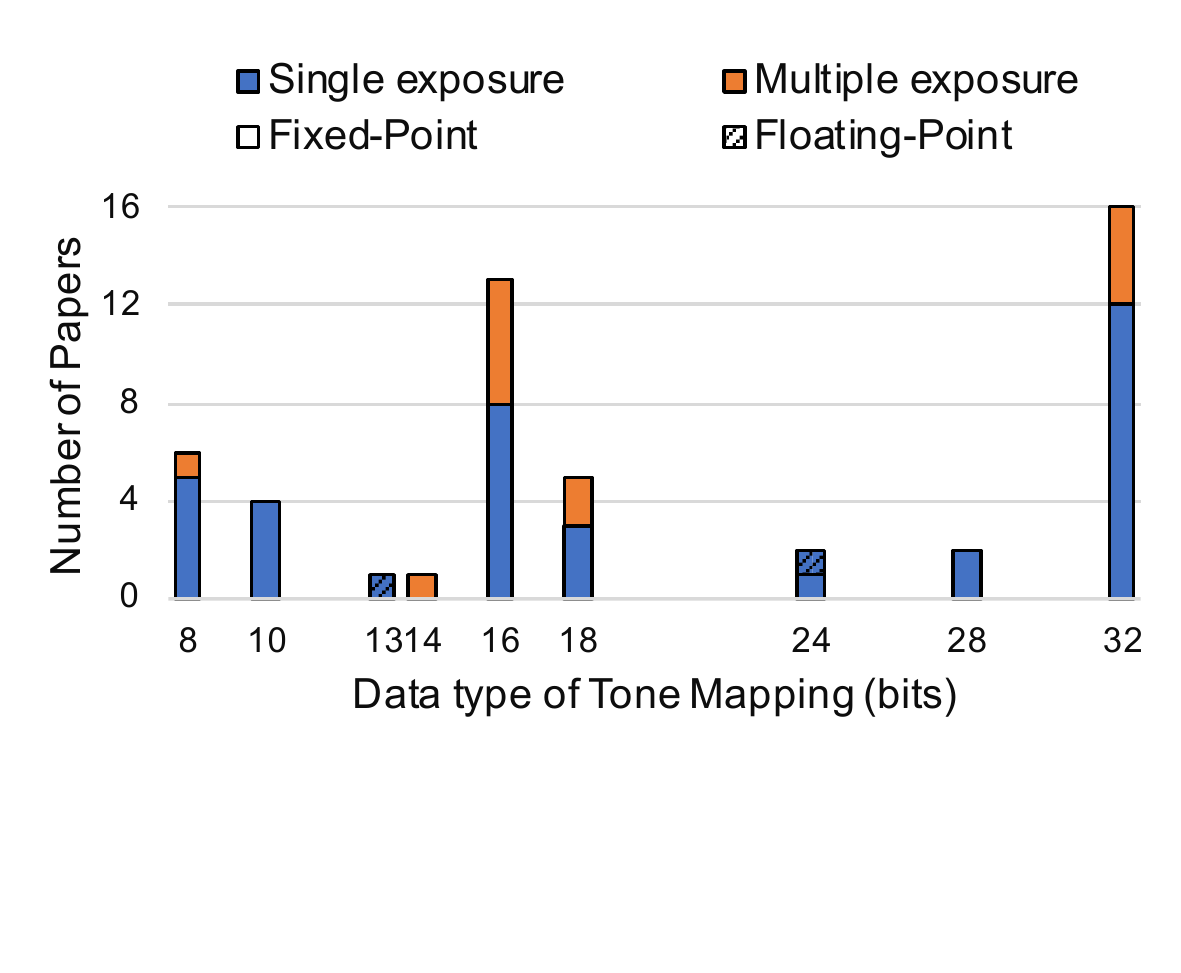}}
\caption{The histogram here shows the numbers of past tonemap operators input data-type. Most of the papers considered 32-bit HDR data.}
\label{fig:dataType1}
\end{figure}

\begin{figure}[!t]
\centerline{\includegraphics[width=8cm, trim=30 0 10 0, clip]{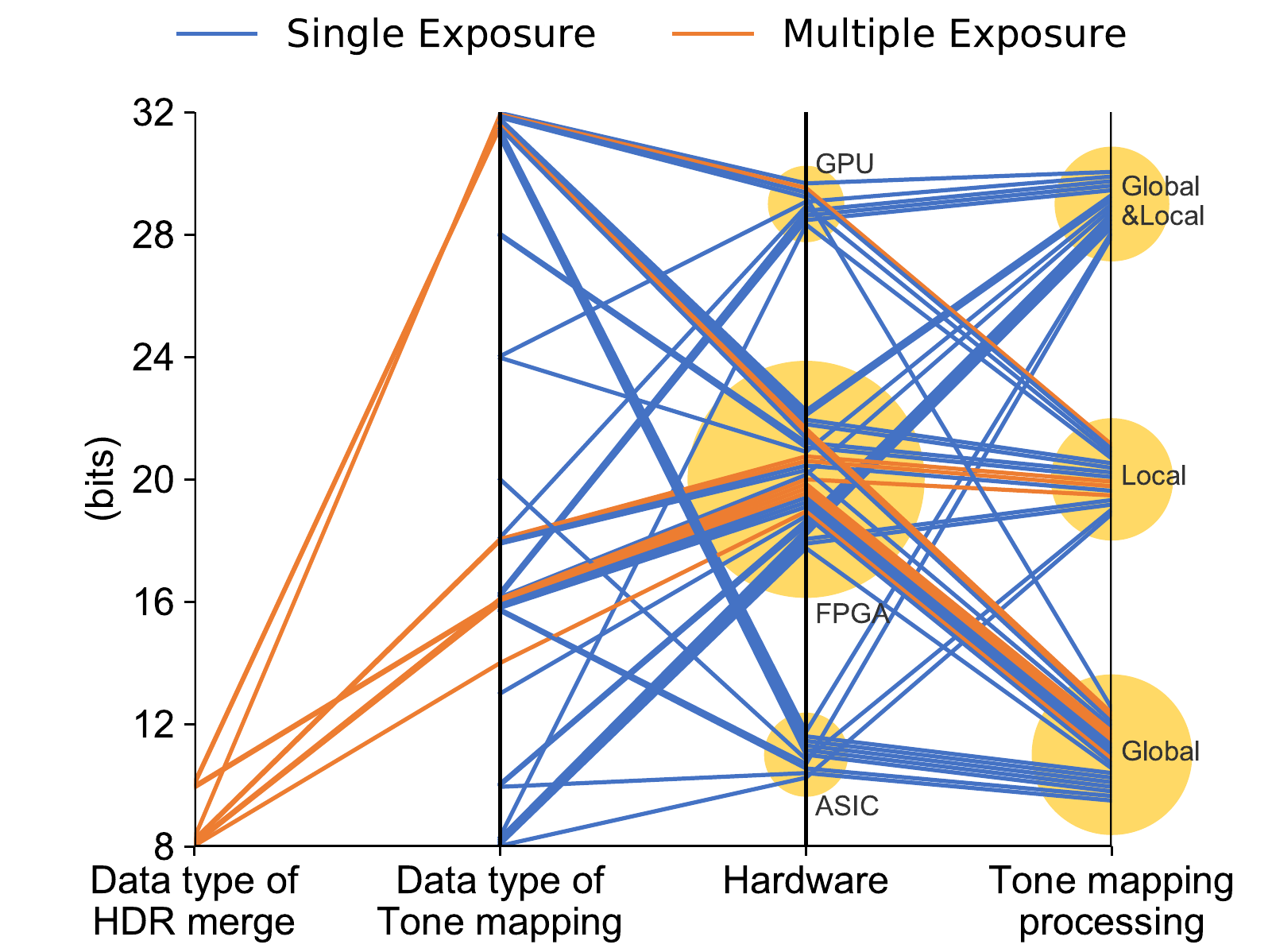}}
\caption{Data type based relationship between HDR images, TMO operators, and hardware platforms.}
\label{fig:dataType2}
\end{figure}

HDR images can be captured from real world scenes, rendered on computers by various computer graphics (CG) tools. In this paper we will focus mainly on methods of obtaining HDR images by using conventional cameras and special HDR sensors, which are useful for building real-time systems. For the CG methods there are well known books describing those methods \cite{ward1998rendering, sillion1994radiosity}. The gaming industry has been employing HDR rendering for very long time, they were used for rendering special visual effects like dazzling, slow dark-adaptation there by enhancing the immersive impression \cite{durand2000interactive}. Today HDR imaging is used in many applications to enhance functionality of cinematography and photography \cite{dufaux2016high}, biomedical imaging (see DICOM standard \cite{mustra2008overview}) \cite{jungmann2011high}, remote sensing \cite{chander2009summary} and many more computer vision applications \cite{szeliski2010computer}. 

Figure \ref{fig:dataType1} shows the frequency of various data types used in all hardware tone mapping papers and whether HDR merge has been implemented in the hardware. Most works choose 16-bit or 32-bit HDR images for tone mapping, and only a few works choose other data types. Almost works proposed TMO on low-power embedded platforms are often implemented using HDR images with fixed-point arithmetic. Compare with floating-point arithmetic, fixed-point arithmetic has some advantages in embedded platforms such as low-power consumption, the small circuit size and high-speed computing \cite{viitanen2013simplified, dobashi2014fixed, lampert2006anisotropic}. On the other hand, due to floating-point arithmetic can save a huge range of luminance with small bit depth. Yadid-Pecht Orly et al. in \cite{ofili2013hardware, yang2019mantissa} proposed a FPGA implementation of tone mapping algorithm for HDR images with floating-point arithmetic.

Different researches implemented different TMO of HDR images with different data type on different hardware. Figure \ref{fig:dataType2} shows how the work of hardware tone-mapping has been distributed and the relationship between data type, hardware and tone mapping processing. 

\section{HDR Merge}

\begin{figure*}[!t]
	\includegraphics [width=\textwidth]{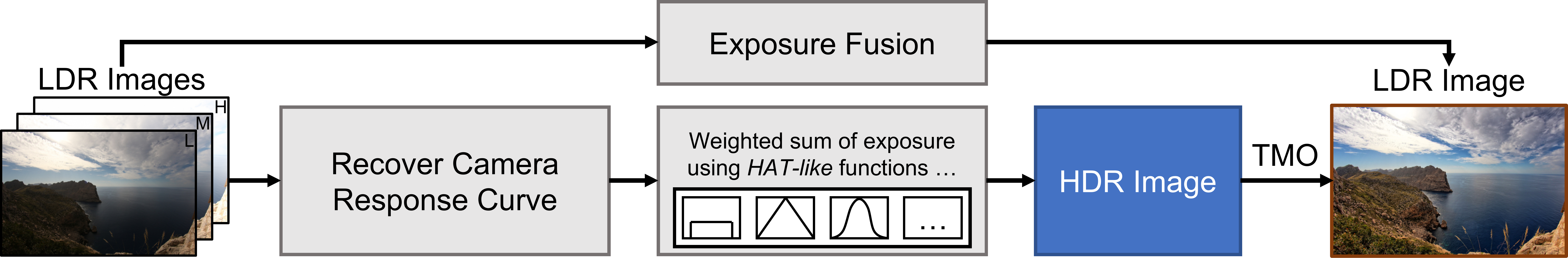}
	\caption{HDR Imaging: Using a camera response curve the full dynamic range of the scene is captured from a set of LDR images with different exposure times. Algorithms like Popadic et al's can directly generate HDR-like images from bracketed images \cite{popadic2017method}. (images from \cite{easyHDR}).}
	\label{fig:EF1}
\end{figure*}

HDR images can also be composed by combining multiple LDR with different exposure time in a single HDR. Here, the exposure time which is also known as the shutter speed is the duration of time when the digital sensor inside the camera is permitted to capture light. The amount of light that reaches the film or image sensor is directly proportional to the exposure time. Therefore, a long exposure time image will have an overall greater luminance. It will detect smaller amount of light sources even in darker areas. But the picture might saturate in bright parts of the scene due to a too much of light for the sensor. On the other hand, a short exposure image will record bright parts of the scene but would not been able to register darker light sources. Exposure time values are often referred to as ``stops''. A stop consists in doubling the exposure time (relative to a reference time). +1 stop is doubling, +2 stops is times 4, and -1 is halving the exposure time \cite{ShutterSpeed, telleen2007synthetic}. So, there are many well known techniques to combine multiple images of varying exposures to compose a HDR image \cite{mannbeing, mitsunaga1999radiometric, debevec2008recovering, tocci2011versatile}.

For an LDR image, its dynamic range is bounded by the range of sensor i.e., a camera with 8 bits/pixel can only capture ratios up to 255:1. However, we can achieve a greater dynamic range using the same camera by combining multiple images which have been captured with different exposure time. Each of these LDR images will cover a different range of the luminance in the scene. This allows to have a greater resolution on the luminance captured . Images with a short exposure time will be adapted for capturing very bright parts of the scene but will fail to capture darker parts. Long exposure time images being the opposite. They will saturate in bright parts of the image. All intermediate images captured with various exposure time will help cover the whole range of luminance, this strategy is demonstrated in Fig. \ref{fig:EF1}. The final composite image will have a greater dynamic range than it is achievable with a single shot by the camera.

\section{HDR Image Sensor}

\begin{figure}[!t]
\centerline{\includegraphics[width=9cm, trim=10 0 0 0, clip]{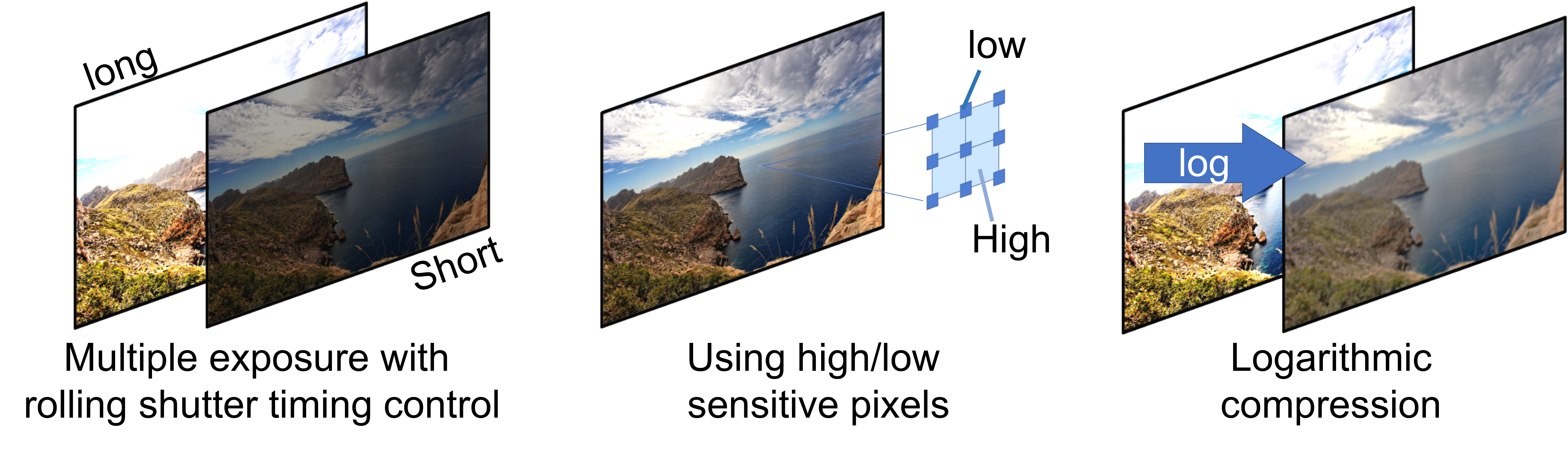}}
	\caption{HDR image sensing methodologies (images from \cite{easyHDR}).}
	\label{fig:HDRsensor}
\end{figure}

There are broadly three different architectures that are used to design HDR image sensors (Fig \ref{fig:HDRsensor}). In the first group, by utilizing logarithmic response pixels or by employing likewise circuits to non-linearly extend the dynamic range as in \cite{kavadias2000logarithmic, loose2001self}. However, this non-linearity can cause severe problems when reconstructing images. The next group sensors like in \cite{sugawa2005100, akahane2006sensitivity, ide2008wide} extend dynamic range by applying lateral overflow capacitors. But, this group of sensors would need to partially open a transfer gate so that the over-saturated charges can be collected by the overflow capacitor. Again, the threshold voltage of the transfer gates can have a large variation, thereby resulting in variations in saturation level. Also, this group of sensors is known to have higher dark current shot noise \cite{liu2012design}. Mase et al., in \cite{mase2005wide} proposed a sensor design where they would use multiple exposure-time to expand the dynamic range. However, this method also has some issues; different integration time can cause discontinuities in SNR and also cause distortion in moving scenes. A new type of HDR image sensor was designed by Liu et al. \cite{liu2012design} which used dual transfer gates.  Their HDR image sensor is capable of capturing HDR scenes with low noise. The novelty of this design is it does not use the concept of transfer of threshold voltage and completes charge transfer in one operation.

\chapter{Tone Mapping Survey}

\section{Tone Mapping General Pipeline}

\begin{figure}[!b]
\centerline{\includegraphics[width=7cm, trim=0 0 -30 0]{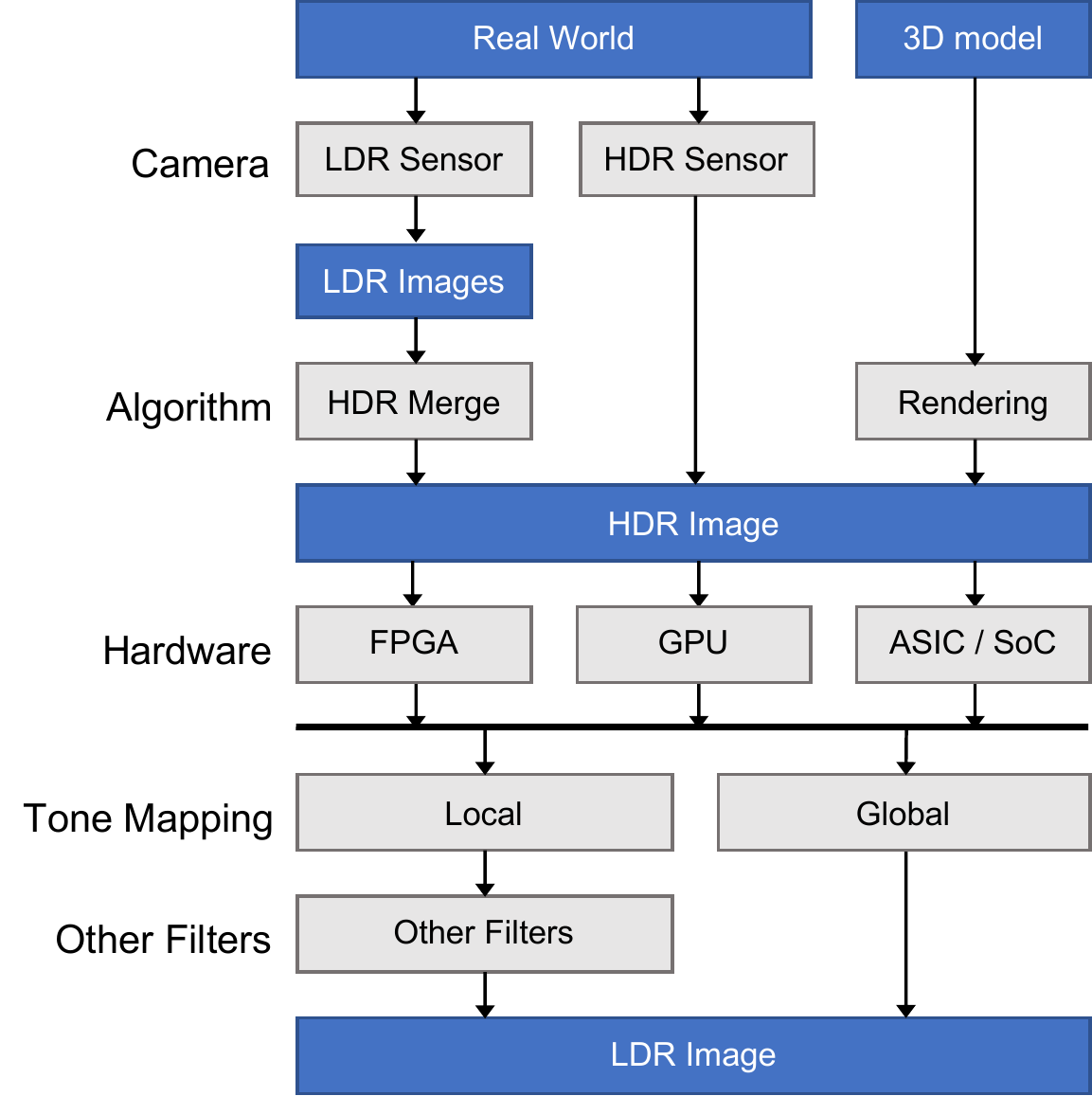}}
\caption{HDR-LDR Pipeline: HDR imaging, sensing, and tone mapping for display. }
\label{fig:generalPipline}
\end{figure}
In the previous section \ref{HDRI} we discussed how to obtain/produce HDR content, and Fig \ref{fig:generalPipline} we show the broad HDR to LDR pipeline. This HDR content requires to be stored in a medium with a greater amount of bits/pixels than that of a single LDR image. It is necessary in order to achieve a greater dynamic range. Although HDR display systems do exist \cite{seetzen2004high}, and TVs with an extended dynamic range are currently available in the commercial market, but they are not as widespread due to their limitations in terms of dynamic range and color gamut. The process of tone mapping consists of reducing the dynamic range of the HDR image into an image that can be easily displayed on wide range of display devices which have limited dynamic range and color gamut. 

This process can be highly non linear depending on the result expected. The LDR image after tone mapping will have less information than the original HDR image due to the reduction of information by the tone map function. But it is ready to be displayed, and the image has enough information by revealing details in both dark and bright parts according to the our perception. Tone mapping has been an important area of research as is evident from several surveys that have been published over the years \cite{wilkie2002tone, dufaux2016high, hoefflinger2007high, myszkowski2008high, reinhard2010high}. Tone mapping function can be classified in to two broad groups based on the processing function they use.

\begin{itemize}
\item Global operators: The same mapping function is applied to all pixels through out image. 
\item Local operators: The mapping function applied to a pixel depends on its neighbors pixels. 
\end{itemize}

\begin{figure}[!h]
\centerline{\includegraphics[width=10cm, trim=0 10 0 0, clip]{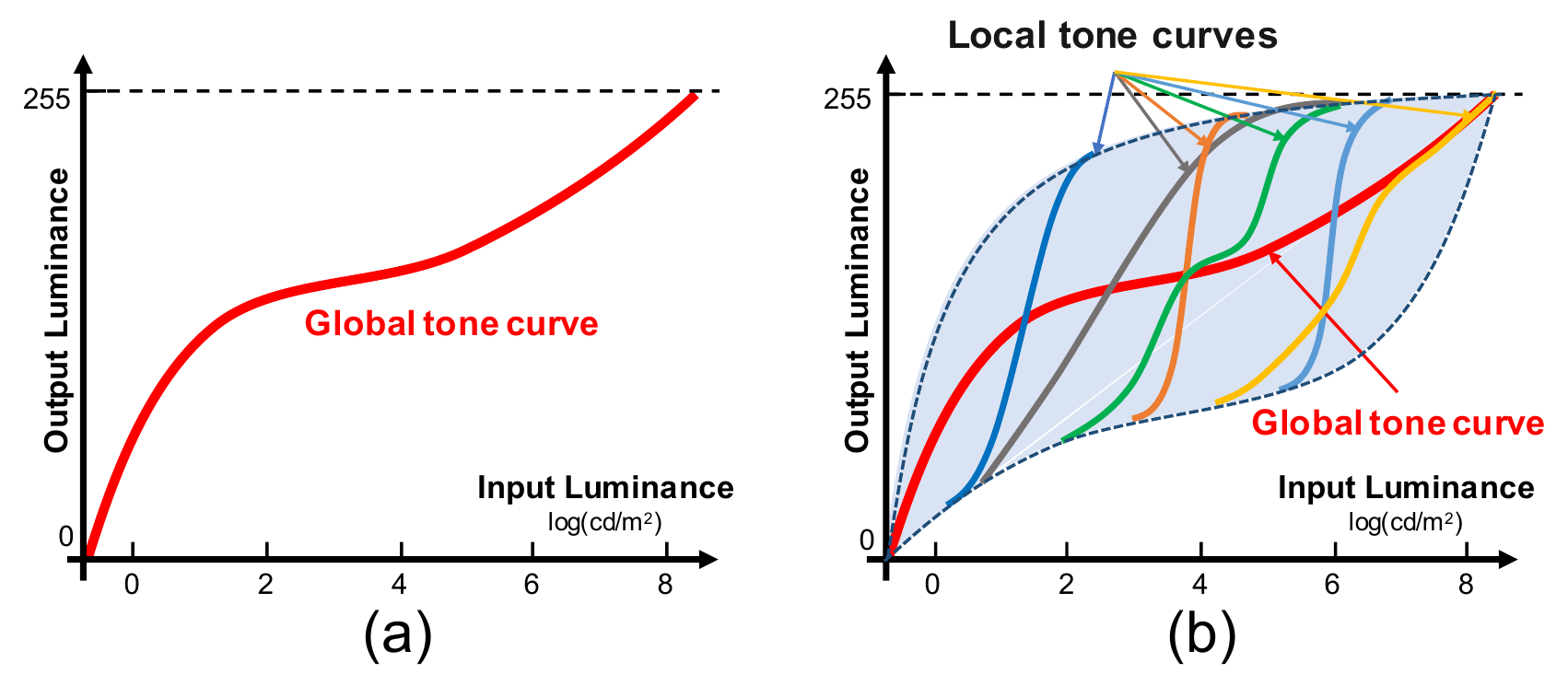}}
\caption{Deformation of tone mapping space. (a) Global Tone Mapping. (b) Global and Local Tone Mapping.}
\label{fig:deformation}
\end{figure}

\begin{figure*}[!t]
	\includegraphics [width=\textwidth, trim=0 5 0 0, clip]{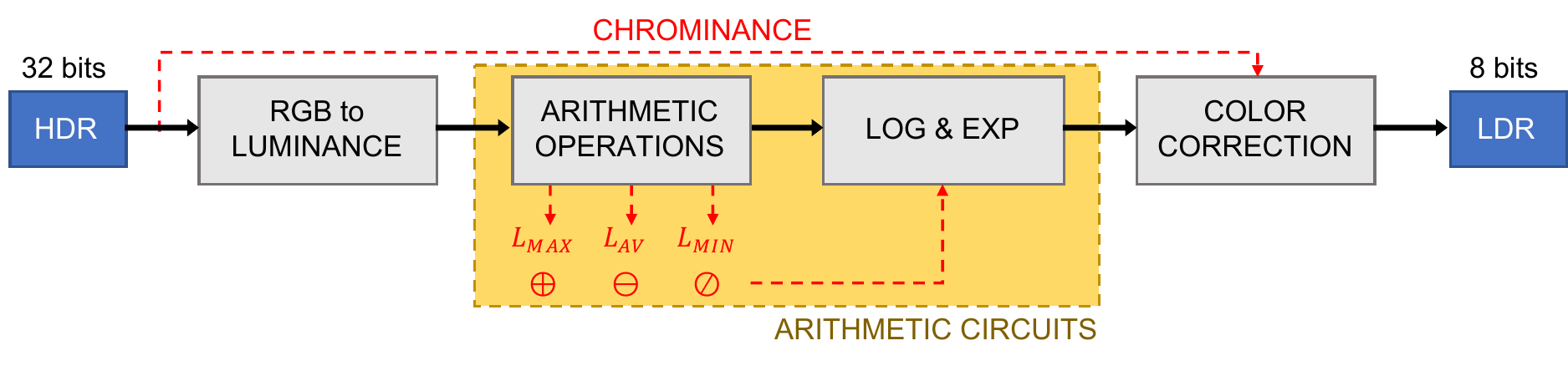}
	\caption{General block diagram for global tone mapping system for hardware implementation.}
	\label{fig:globalPipline}
\end{figure*}

\section{Global Tone Mapping Algorithms} 

Global tone mapping algorithms (also known as ``tone reproduction curves'') are spatially invariant, that is they apply the same function to all pixels in the input image \cite{ward1994contrast,schlick1995quantization,ferwerda1996model,tumblin1999two,drago2003adaptive}. This results in one and only one output value for a particular input pixel value irrespective of the pixels in its neighborhood. Figure \ref{fig:globalPipline} shows a general pipeline which is useful for implementing a global tonemap function shown in Fig. \ref{fig:deformation} (a). As we can see, the tonemap pipeline first obtains the luminance image and from that it calculates global statistics (like $L_{max}, L_{min},L_{average}$). In some algorithms these statistics are also calculated from previous frame based on a assumption that there is very little changes between successive frames when imaging at 30/60 frames per second \cite{lapray2011smart,lapray2012hdr,popovic2014performance}.

In the pipeline next step is to realize a logarithmic or exponential like function to compute the tone mapped image. The hardware implementation of these functions are a challenge and a common approach is to approximate such functions, by maintaining an acceptable level of error in the realized implementation. As these functions are frequently required to be computed in many engineering and scientific applications many hardware friendly methods have been  proposed in literature \cite{kantabutra1996hardware,koren2001computer,parhami2010computer}. The final step in the pipeline after tonemap is to restore the color for displaying the output image. A detailed account for color correction technique is presented in section \ref{c2l}.

\begin{figure*}[!t]
	\includegraphics [width=\textwidth]{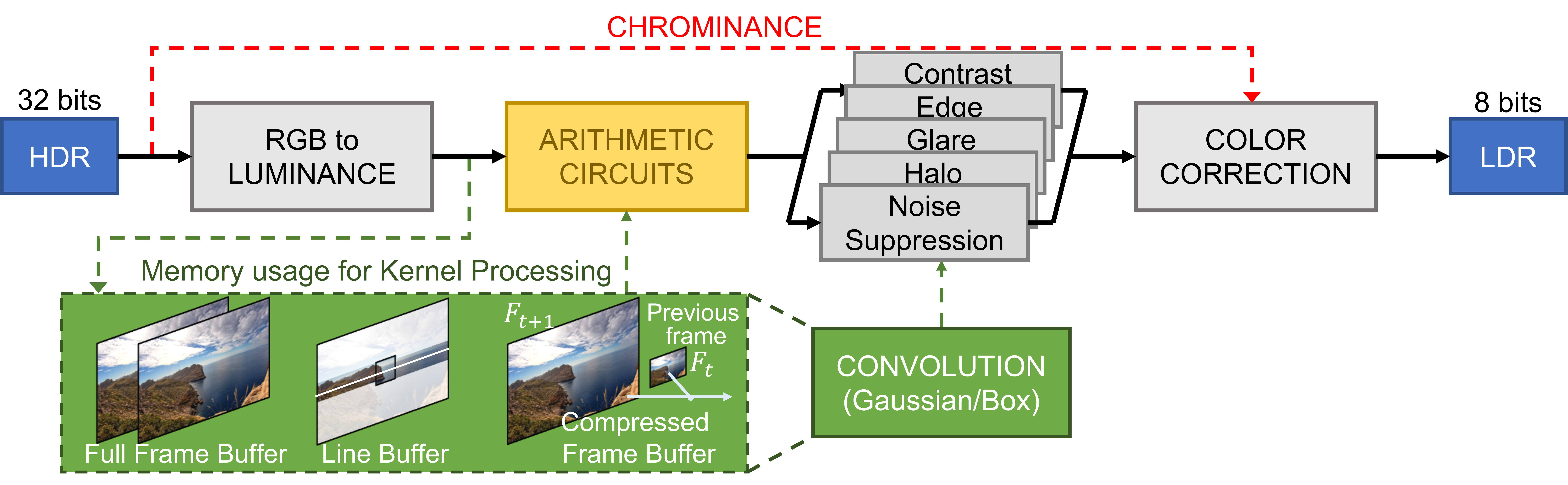}
	\caption{General block diagram for local tone mapping system for hardware implementation. For local calculation full frame buffer/line buffer or compressed frame buffer is required (images from \cite{easyHDR}).}
	\label{fig:localPipline}
\end{figure*}

\section{Local Tone Mapping Algorithms}

Local tone mapping algorithms (also known as 'tone reproduction operators') are spatially variant and apply different functions based on the neighborhood of the input pixel \cite{chiu1993spatially, pattanaik2000time, durand2002fast, reinhard2002photographic, ashikhmin2002tone, pattanaik2002adaptive, colbert2007painting}. Therefore, one input pixel could result in different output values based on its position as illustrated in Fig. \ref{fig:deformation} (b). Local tone mapping algorithms are computationally more expensive and time consuming compared to global tone mapping algorithms \cite{wang2007design}. We can describe the operation of a local tonemap algorithm using the block diagram shown in Fig \ref{fig:localPipline}. As was in the case of global tonemap, we initially obtain the luminance values for the input image. The color correction step is described in detail in section \ref{c2l}. The high computation cost for local tonemap operator is due to the local information calculation for which a full frame or a few lines of the input image has to be buffered as shown in the Fig \ref{fig:localPipline}. Some algorithms have also implemented compressed frame buffer (down-sampled images) to reduce the memory cost \cite{marsi2007video, ambalathankandy2019adaptive}. To meet the real-time constraints, as a common approach previous frame is used to compute the local information for current frame.

Local tone mapping methods generally produce better quality images as they preserve details, which may not be the case when using global tone mapping methods \cite{reinhard2010high}. However, one of the major drawbacks of local tone mapping algorithms are the creation of halo artifact among the high contrast edges and the graying out of the low-contrast areas \cite{herscovitz2004modified,meylan2006high}. Therefore, local tonemap methods implement additional filters to suppress these image artifacts like halo and noise. Such filtering may require that the input image (of size $M \times N$) be convolved with a filter (of size $k \times r$). Benedetti et al. demonstrated a simple hardware sliding window convolution block, which can output one pixel every clock \cite{benedetti1998image}. The latency associated with this sliding window method is calculated as:

\begin{equation}
T = Buffer_{depth} \times \floor*{\frac{Kernel Size}{2}} + \ceil*{\frac{Kernel Size}{2}}   
 \label{Eq:3} 
\end{equation}

\section{Color to Luminance}\label{c2l}

Image luminance are used for various applications such as printing, data compression, feature extraction and tone mapping. Thus, obtaining luminance is a very important step for TMOs. There are many well defined methods to obtain the luminance values from the color image. An easy method to obtain luminance is to compute it as a linear combination of the red, green, and blue component according to the RGB-to-XYZ conversion scheme. Here, 

\begin{equation}
Y = 0.2126R + 0.7152G + 0.0722B
\label{Eq:1}
\end{equation}

is the luminance for the RGB image. Another effortless procedure is to use CIELab or YUV color spaces, and one could directly obtain luminance channel as the grayscale version of the color image. Because, they consider the luminance and color channel to be independent.  

Tone mapping algorithms operate on the luminance channel which are obtained as described above. For a given HDR image luminance value is calculated and the chrominance values are buffered/stored as they are required later for restoring the color post tonemap. Different studies have used different luminance methods, which we have listed in table \ref{table_1_1}. Some studies have also used monochrome images \cite{hassan2007fpga,lapray2011smart,li2016novel,vargas2014151,shi2016tone}, this approach can have certain advantages in terms of reduced memory, and fewer calculations. However, the application of monochrome images are limited. 

After tone mapping, a common approach to restore the color is based on Schlick's color ratios \cite{schlick1995quantization}:

\begin{equation}
C_{out} = \big(\frac{C_{in}}{L_{in}}\big)^{\gamma}L_{out}
\label{Eq:2}
\end{equation}

In Eq. \ref{Eq:2}, $C_{in}$ represent the original RGB image, $L_{in}$ is the corresponding luminance value obtained by Eq. \ref{Eq:1}. If $L_{out}$ is the tone mapped luminance value then, we can compute three output chrominance values as in Eq. \ref{Eq:2}, where $\gamma$ is a color saturation factor for displaying color images, and its value is usually set between 0.4 and 0.6 \cite{fattal2002gradient}.
\chapter{Tone Mapping Operators Implemented on Hardware Platform}\label{sec4}

\begin{figure}[!h]
\centerline{\includegraphics[width=9cm, trim=0 30 -20 10, clip]{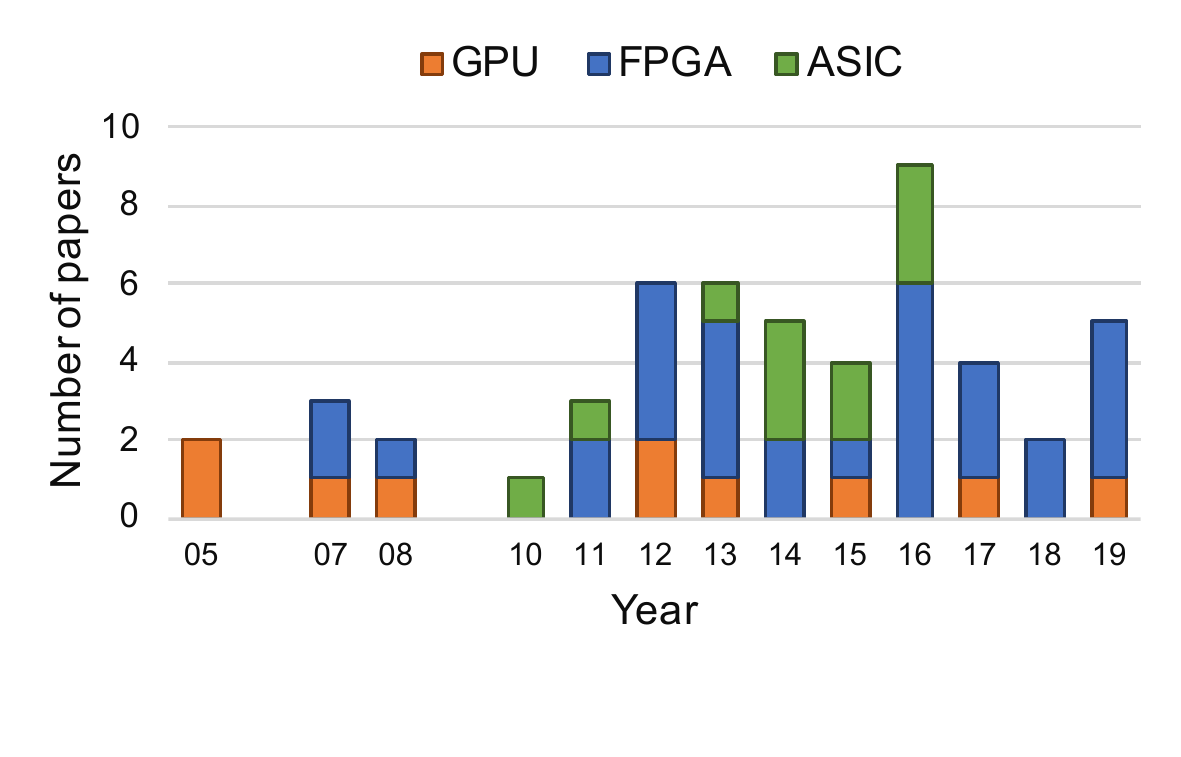}}
\caption{The number of TMO hardware implementation papers in the last 15 years.}
\label{fig:paperNumber}
\end{figure}

Tone mapping HDR images/videos is a computationally challenging problem, and many hardware-based dedicate accelerators have been proposed for this purpose. In this section we will discuss those systems in detail. We have comprehensively listed these implementations in table \ref{table_1_1}, and summarized the number of these papers by year in Fig \ref{fig:paperNumber}.

\newcommand{\tabincell}[2]{\begin{tabular}{@{}#1@{}}#2\end{tabular}}

\begin{table*}\tiny
	\renewcommand{\arraystretch}{1.4}
	\caption{LIST OF TMO HARDWARE IMPLEMENTATIONS}
	\centering
	\label{table_1_1}
	\resizebox{\textwidth}{!}{
    \setlength{\tabcolsep}{0mm}{
    \begin{tabular}{|ccc|c|ccc|cccc|ccc|}
    \hline
    \multicolumn{3}{|c|}{} &  & \multicolumn{3}{c|}{\textbf{TMO}} & \multicolumn{4}{c|}{\textbf{Image}} & \multicolumn{3}{c|}{\textbf{Image Quality}} \\
    \hline
    \textbf{} & \tabincell{c}{Previous\\work} & Year & \tabincell{c}{Hard-\\ware} & Method & Reference & \tabincell{c}{Kernel\\Size$^\epsilon$} & Channel & \tabincell{c}{Lumi-\\nance} & \tabincell{c}{Data type\\(bit)} & \tabincell{c}{HDR\\Merge} & \tabincell{c}{PSNR\\(dB)} & SSIM & Others \\
    \hline\cite{goodnight2005interactive}&       & 2005  & GPU   & G\&L$^\gamma$ &\cite{reinhard2002photographic} & $49\times49^{8\zeta}$ & Color &  & 24    & No    &       &       & RMS\%=1.051  \\
    \hline\cite{krawczyk2005perceptual}&       & 2005  & GPU   & Local &\cite{reinhard2002photographic}$^\delta$ & $61\times61^{8\zeta}$ & Color & XYZ & 8to32$^\kappa$ &\cite{mantiuk2004perception} &       &       & RMS\%=0.245  \\
    \hline\cite{roch2007interactive}&       & 2007  & GPU   & Local &\cite{ashikhmin2002tone} & $5\times5^{2\zeta}$ & Color &  & 32    & No    &       &       & RMS\%=0.038 \\
    \hline\cite{zhao2008real}&       & 2008  & GPU   & - &\cite{drago2003adaptive, reinhard2002photographic, durand2002fast, ashikhmin2002tone, pattanaik2000time, pattanaik2002adaptive, colbert2007painting} & - & Color &  & 32    & No    &       &       &  \\
    \hline\cite{tiant2012gpu}&       & 2012  & GPU   & Local &\cite{duan2010tone}$^\delta$ & $64\times64$ & Color &  & 18 & No    &       &       &  \\
    \hline\cite{akil2012real}&       & 2012  & GPU   & G\&L$^\gamma$ &\cite{irawan2005perceptually} & - & Color & XYZ & 32 & No    &       &       &  \\
    \hline\cite{urena2013real}&\cite{urena2012real}& 2013  & GPU   & G\&L & - & $11\times11$ & Color & HSV & 16 & No    &       &       & \tabincell{c}{AMBE=108.09 DE=4.62\\EME=11.306}  \\
    \hline\cite{eilertsen2015real}&  & 2015  & GPU   & G\&L$^\gamma$ & - & $230\times230$ & Color &  & 16 & No    &       &       &  \\
    \hline\cite{khan2017tone}&       & 2017  & GPU   & Global & \cite{larson1997visibility}$^\delta$ & N/A   & Color &  & 32    & No    &       &       & TMQI=0.9236 \\
    \hline\cite{tsai2019real}&\cite{li2015low}  & 2019  & GPU   & Local &\cite{tsai2011novel} &  multiscale$^\zeta$ & Color &  & 8     & No    &       &       & \\ 
    \hline\cite{hassan2007fpga} &  & 2007 & FPGA & Local &\cite{reinhard2002photographic} & $64\times64^{9\zeta}$ & Gray & $-^\eta$ & 28    & No    & 34.94 & &   \\
    \hline\cite{marsi2007video} &  & 2007 & FPGA & G\&L$^\gamma$ &\cite{marsi2004image} & $5\times5$ & Color & HSV & 10    & No    &  & &   \\
    \hline\cite{iakovidou2008fpga} &  & 2008 & FPGA  & G\&L$^\gamma$ &\cite{vonikakis2008fast} & $61\times61$ & Color & YCbCr & 8 & No    & & &   \\
    \hline\cite{vakili2011customized} &  & 2011 & \tabincell{c}{FPGA\\ASIP$^\alpha$}  & Global &\cite{reinhard2002photographic} & N/A & Color &  & 10.22$^\lambda$ & No & 52.95 & &   \\
    \hline\cite{lapray2012hdr} &\cite{lapray2011smart}  & 2012 & FPGA  & Global &\cite{duan2010tone} & N/A & Gray &  & 10to16$^\kappa$ &\cite{debevec2008recovering} & & & \\
    \hline\cite{kiser2012real} &  & 2012 & FPGA  & Global &\cite{reinhard2002photographic} & N/A & Color & $-^\eta$ & 18    & No    &  & &   \\
    \hline\cite{urena2012real} &  & 2012 & \tabincell{c}{FPGA\\GPU$^\beta$}  & G\&L$^\gamma$ &\cite{tsai2011novel, tsai2015adaptive} & $7\times7$ & Color & HSV   & 16  & No    & 17.35 & &   \\
    \hline\cite{mann2012realtime} &  & 2012 & FPGA  & Global &\cite{kang2003high, pal2004probability, granados2010optimal} & N/A & Color &  & 8to14$^\kappa$ &\cite{robertson2003estimation, ali2012comparametric, mann1993compositing} & & &  \\
    \hline\cite{lapray20131} &\cite{lapray2012hdr} & 2013 & FPGA  & Global &\cite{duan2010tone, reinhard2002photographic} & N/A & Gray &  & 10to32$^\kappa$ &\cite{debevec2008recovering} &  &  &  \\
    \hline\cite{ofili2013hardware} &  & 2013 & FPGA  & G\&L$^\gamma$ &\cite{ofili2012depth} & $3\times3$ & Color &  & 20.12$^\lambda$ & No & 54.27 & 0.9997 & \tabincell{c}{RMSE=0.932 SSE=0.233\\R-squared=0.03549} \\
    \hline\cite{vytla2013real} &\cite{hassan2007fpga} & 2013 & FPGA & Local &\cite{fattal2002gradient} & $3\times3$ & Gray & $-^\theta$ & 32 & No    & 73.96 & &   \\
    \hline\cite{canada2013embedded} &  & 2013 & FPGA & G\&L$^\gamma$ &\cite{pizer1987adaptive} & $7\times7$ & Color & HSI & 8 & No & 30 & &   \\
    \hline\cite{popovic2014performance} &  & 2014 & FPGA  & Global &\cite{drago2003adaptive} & N/A & Color & YUV   & 16    & No    & 60.03 & 0.9995 &   \\
    \hline\cite{shiau2014low} &  & 2014 & \tabincell{c}{FPGA\\SoC$^\beta$}  & Local &\cite{bhuiyan2008fast} & $3\times3$ & Color & HSV   & 8     & No    & & & \tabincell{c}{DE=5.01 EBCM=0.032\\CEF=1.6}   \\
    \hline\cite{li2015low} &  & 2015 & FPGA  & Local &\cite{tsai2012fast} & $32\times32^{3\zeta}$ & Gray & HSV & 8 & No & 43.37 & 0.9982 & $\Delta E_{HS}$=0.0039   \\
    \hline\cite{lapray2016hdr} &\cite{lapray20131} & 2016 & FPGA  & Global &\cite{duan2010tone, reinhard2002photographic} & N/A & Gray &  & 10to32$^\kappa$ &\cite{debevec2008recovering} & 23.55 & 0.93 & MSE=286.95 UQI=0.9 \\ 
    \hline\cite{ambalathankandy2019fpga}  &  & 2016 & FPGA  & G\&L$^\gamma$ &\cite{hore2014statistical, hore2014new} & $5\times5$ & Color &  & 20.12$^\lambda$ & No    & 57.27 & 0.9969 &  \\
    \hline\cite{shahnovich2016hardware} &  & 2016 & FPGA  & G\&L$^\gamma$ & - & $3\times3$ & Color &  & $16^8$ $^\mu$ & No    & 55.87 & 0.9996 &   \\
    \hline\cite{liu2016study} &\cite{vytla2013real} & 2016 & FPGA  & Local &\cite{fattal2002gradient} & $7\times7$ & Gray &  & 28    & No    & & & GCC=0.75 DM\%=0.88   \\
    \hline\cite{popovic2016multi} &\cite{popovic2014performance} & 2016 & FPGA  & Global &\cite{drago2003adaptive} & N/A & Color & YUV   & 8to16$^\kappa$ &\cite{mertens2007exposure} & 103.61 & &   \\
    \hline\cite{li2016novel} &  & 2016 & \tabincell{c}{FPGA\\CPU$^\alpha$}  & Global & - & N/A & Color & HSV & 16 & No &  &  &   \\
    \hline\cite{nosko2017true} &  & 2017 & FPGA  & Local &\cite{durand2002fast} & $9\times9$ & Color & $-^\theta$ & 8to16$^\kappa$ &\cite{debevec2008recovering} & 39 & &   \\
    \hline\cite{zemvcik2017real} &  & 2017 & FPGA  & Local &\cite{banterle2008hdr} & $11\times11$ & Color & YCbCr & 8to18$^\kappa$ &\cite{myszkowski2008high} &  & &   \\
    \hline\cite{popadic2017method} &  & 2017 & FPGA  & Global & - & N/A & Color &  & 8to8$^\kappa$ & Yes &  & &   \\
    \hline\cite{nosko2018color} &\cite{nosko2017true} & 2018 & FPGA  & Local &\cite{durand2002fast} & $9\times9$ & Color & $-^\theta$ & 10.8$^\lambda$ &\cite{debevec2008recovering, grosch2006fast} & 46.69 & 0.93 &   \\
    \hline\cite{yang2018local} &  & 2018 & FPGA  & Local & - & $32\times32$ & Color &  & 20.12$^\lambda$ & No    & & & TMQI=0.9266  \\
    \hline\cite{yang2019mantissa}&\cite{shahnovich2016hardware}& 2019 & FPGA  & Global & - & N/A & Color &  & $10^3$ $^\mu$ & No    & & & TMQI=0.9111  \\
    \hline\cite{liu2019high} &  & 2019 & FPGA  & Global &\cite{zhang2016retina} & N/A & Color &  & 16 & No    & 81.47 & 0.9998 &   \\
    \hline\cite{ambalathankandy2019adaptive} &\cite{ambalathankandy2019fpga} & 2019 & FPGA  & G\&L$^\gamma$ & \cite{shimoyama2009local, igarashi2013accuracy, kimura2015halo} & $31\times31$ & Color &  & 10 & No & & 0.9696 & TMQI=0.9360   \\
    \hline\cite{park2019low} &  & 2019 & FPGA  & Local &\cite{shin2015efficient} & $29\times29$ & Color &  & 5.13$^\lambda$  & No    & & &   \\
    \hline\cite{chiu2011real}& & 2010  & ASIC & G\&L$^\gamma$ &\cite{reinhard2002photographic, fattal2002gradient} & $8\times8$ & Color & $-^\iota$ & 16.16$^\lambda$ & No    & 45/35 &       &  \\
    \hline\cite{punchihewa2011review}& & 2011  & SoC & Global &\cite{kagami2007study} & N/A & Gray &  & 16    &\cite{nayar2000high, gu2010coded, sasaki2007wide}    &       &       &  \\
    \hline\cite{sicard2013cmos}& & 2013  & Analog & Global &\cite{meylan2007model} & N/A & Color &  & HDR & No &       &       &  \\
    \hline\cite{vargas2014151}& & 2014  & VSoC & Global &\cite{reinhard2010high} & N/A   & Gray &  & HDR    &\cite{mantiuk2004perception} & 53.8  &       &  \\
    \hline\cite{gouveia2014reconfigurable}& & 2014  & Analog & Local &\cite{drago2003adaptive, reinhard2002photographic} & - & Gray &  & 16    & No    &       &       &  \\
    \hline\cite{mughal2014threshold}&  & 2014  & Analog & - &\cite{schlick1995quantization, banterle2017advanced, drago2003adaptive} & - & Gray &  & 10    & No    &       &       &  \\
    \hline\cite{mughal2015fixed}& \cite{mughal2014threshold} & 2015  & Analog & Global &\cite{reinhard2002photographic} & N/A & Gray &  & 10    & No    &       &       &  \\
    \hline\cite{fernandez2015single}& & 2015  & Analog & Global & - & N/A & Gray &  & HDR    & No    &       &       &  \\
    \hline\cite{shi2016tone}& \cite{shi2016analog} & 2016  & Analog & Global &\cite{reinhard2002photographic} & N/A   & Gray &  & 32    & No    &       &       &  \\
    \hline\cite{chen2016analog}& & 2016  & Analog & G\&L$^\gamma$ &\cite{reinhard2002photographic, duan2011local} & $3\times3$   & Color & YUV   & 16    & No    & 34.45 & 0.987 &  \\
    \hline\cite{guicquero2016algorithm}& & 2016  & Analog & Global &\cite{reinhard2010high} & N/A   & Gray &  & 32    &\cite{spivak2009wide} & 35 &       &  \\
    \hline
    \end{tabular}
    }}
    \tiny
    \\$^\alpha$Implemented on two types hardware together $^\beta$Implemented on two types hardware respectively
    \\$^\gamma$Global and local tone mapping operator $^\delta$An improvement of cited tone mapping operator
    \\$^\epsilon$Local luminance calculation kernel size $^\zeta$($i\times i^k$):$k$ scales Gaussian pyramid with a maximum size of $i\times i$
    \\$^\eta$0.27R+0.67G+0.06B $^\theta$0.299R+0.587G+0.114B $^\iota$0.2654R +0.6704G +0.0642B
    \\$^\kappa$($k$ to $i$):$k$bit LDR images into $i$bit HDR image by HDR merge
    \\$^\lambda$($k.i$):$k$bits integer and $i$bits fraction $^\mu$($k^i$):$k$bits mantissa and $i$bits exponent
\end{table*}

\section{Design Bottlenecks: Software to Hardware Porting}

\begin{figure}[!b]
\centerline{\includegraphics[width=9cm]{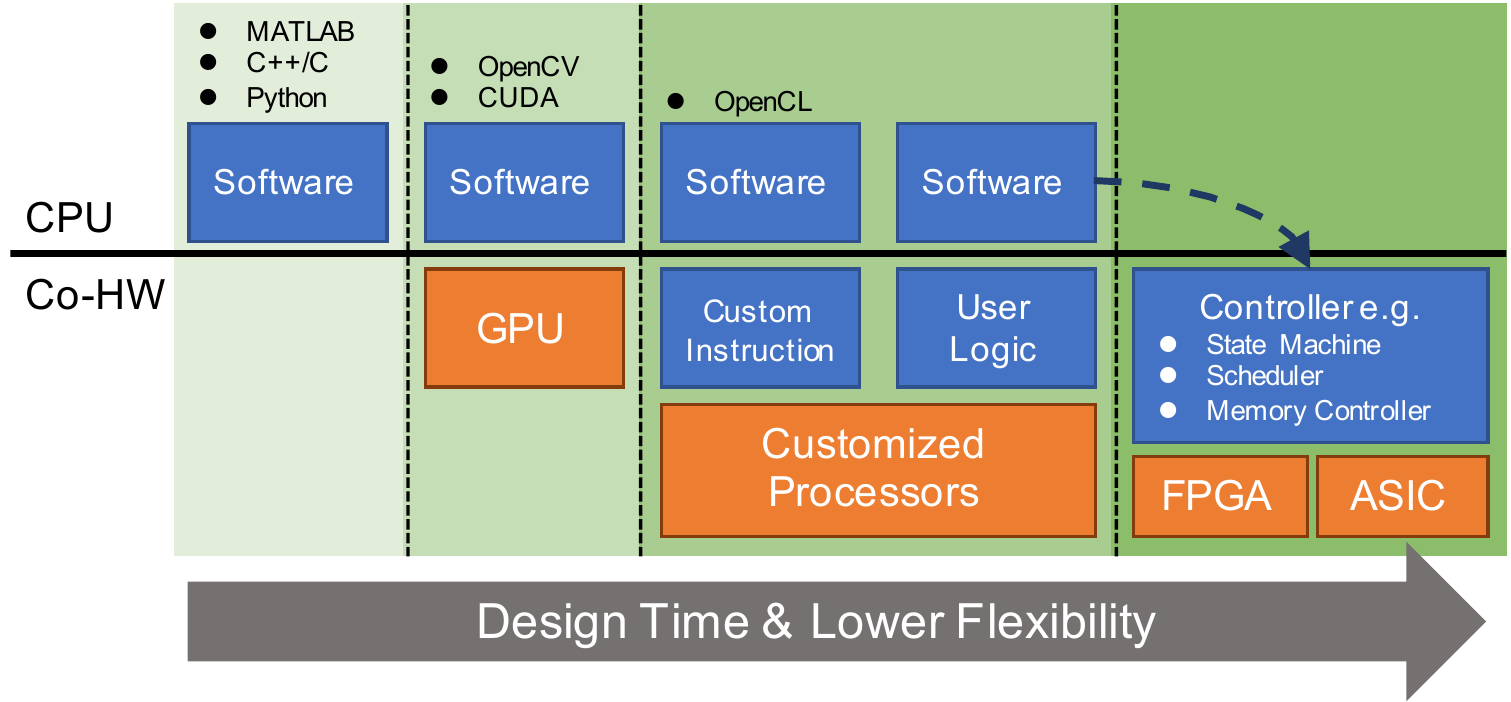}}
\caption{Choice of hardware platform for developing an algorithm mainly depends upon the application. Other important factors are design time and engineering costs.}
\label{fig:HWDesign}
\end{figure}

Real-time image processing applications are bound timing constraints, i.e., for every pixel clock the processing pipeline should consume one pixel from the camera side and deliver one to the display side. Any missed pixel on either side would lead to loss of information or cause blanking display, this is known as the throughput constraint. When porting software algorithms to hardware an inherent design problem is that the SW code is developed on a general purpose CPU. Therefore, the algorithm is highly sequential, and it is useful to exploit the fast CPU. However, this is not the case on HW platform. For example, FPGAs are clocked at much lower frequencies and designers should exploit this parallelism to implement real-time systems. Another type of constraints that has to be met for real-time tone mapping system, is the pipeline latency. Here, latency implies how many clock cycles are required to process one input pixel to processed/tone-mapped pixel.

Memory bottleneck is crucial for implementing image processing algorithms on hardware. While HW platforms like FPGAs have highly parallel logic blocks and fast reconfigurable interconnects to speed up window (kernel) operations. The interface speed between the tonemap accelerator and the rest of the FPGA system is a bottleneck. Particularly, for image processing applications whose data bandwidth requirements are extremely high volume. The cost of moving data between off-chip memory and the accelerator can be detrimental and outweigh the benefits of implementing the FPGA system. Therefore, well thought out operation sequence that obeys raster order should be chosen, because other computation order would usually require the whole frame to be buffered. Like caching can reduce the memory bottlenecks on CPUs, streaming FIFO interfaces can reduce the amount of pixel accesses on FPGA hardware. Also, FPGAs are provided with BRAMs which can be read and written at the same time at every clock cycle, allowing one stream of values to be stored and one stream to be extracted in parallel. 

Various technologies are available for implementing image/video processing algorithms. However, the main concerns of design implementation are cost, speed and power. The design methodology adopted for any hardware implementation depends on the application and time to market. The hardware-based implementation can be realized using any one of these hardware platforms: Application-Specific Integrated Circuits (ASIC), Field-Programmable Gate Arrays (FPGA), and Graphics Processing Units (GPU). Each have its own advantages and disadvantages, which we will briefly explain using Fig. \ref{fig:HWDesign}. From this figure, we can notice that the choice of platform depends on various factors like: flexibility, design time and cost. A full custom ASIC design development will be very expensive due to increased manufacturing and design time, increased non-recurring engineering costs. Even though the ASIC design solution can be very efficient in terms of area and performance, it is only viable for proven designs that can be mass produced. GPUs and FPGA platforms have been preferred for many image processing applications and we will discuss more about them in the following sections \ref{GPU} and \ref{FPGA}. OpenCL-based is suitable for implementing algorithms on general-purpose computing on graphics processing units (GPGPU), and it has flavor similar to the proprietary CUDA language from NVIDIA. Recently, FPGA vendors are also supporting openCL development processes. 

\section{Graphics Processing Unit}\label{GPU}
GPUs came to prominence in 1990s to support more graphically realistic computer games, as it became hard to support good graphics and performance using CPUs only. GPUs are efficient and lot faster than a CPU in terms of floating point operations per second as they are specially developed for compute-intensive and highly parallel computations. As we know that image processing tasks are well-suited for parallel computing, and an average image consists of millions of individual pixels is a good case for GPU processing.
Table \ref{table_2_1} lists TMOs implemented on GPUs, and will be discussed in this section.  
\begin{table}[!b]\scriptsize
  \renewcommand{\arraystretch}{1.2}
  \caption{GPU IMPLEMENTATIONS}
  \centering
  \label{table_2_1}%
    \setlength{\tabcolsep}{1mm}{
    \begin{tabular}{|c|cc|ccc|}
    \hline\multicolumn{1}{|c|}{} & \multicolumn{2}{c|}{\textbf{Hardware}} & \multicolumn{3}{c|}{\textbf{Performance}} \\
    \hline\textbf{}  & GPU & \tabincell{c}{Technology\\(nm)} & \tabincell{c}{Frame Size\\(pixel)} & \tabincell{c}{Speed\\(FPS)} & \tabincell{c}{Throughput\\(Mpix/s)} \\
    \hline\cite{goodnight2005interactive}& Radeon 9800 Pro & 150 & $512\times512$ & 5 & 1.3 \\
    \hline\cite{krawczyk2005perceptual}& GeForce 6800GT & 130 & $1024\times768$ & 10 & 7.9 \\
    \hline\cite{roch2007interactive}& GeForce Go 6800 & 130 & $2048\times2048$ & 7 & 29.4 \\
    \hline\cite{zhao2008real}& GeForce 8800 GTS & 90 & - & - & - \\
    \hline\cite{urena2012real}& GeForce 7900 GTX & 90 & $640\times480$ & 30 & 9.2 \\
    \hline\cite{tiant2012gpu}& GeForce GT 550M$\times2$ & 40 & $1024\times768$ & 2.8 & 2.2 \\
    \hline\cite{akil2012real}& GeForce 8800 GTX & 90 & $1002\times666$ & 37 & 24.7 \\
    \hline\cite{urena2013real}& NVIDIA ION2 & 40 & $640\times480$ & 27 & 8.3 \\
    \hline\cite{eilertsen2015real}& GeForce GTX 980 & 28 & $1980\times1080$ & 46.5 & 99.4 \\
    \hline\cite{khan2017tone}& GeForce Titan Black & 28 & $2048\times1536$ & 24 & 75.5 \\
    \hline\cite{tsai2019real}& GeForce GTX 650 Ti & 28 & $4096\times4096$ & 7.5 & 125.8 \\
    \hline
    \end{tabular}%
    }
\end{table}%

Goodnight et al. \cite{goodnight2005interactive} proposed a tone mapping algorithm implementation \cite{reinhard2002photographic} using programmable graphics hardware. This work also discussed, some applications of tone mapping for rendering. They cleanly map the tone mapping algorithm to the pixel processor, which allows an interactive application to achieve higher levels of realism by rendering. They also describe how the graphics hardware limits the effective compression of dynamic range and discuss modifications to the algorithm that could alleviate these problems. 

Krawczyk et al. \cite{krawczyk2005perceptual} propose a model include HDR image perception effects by local TMO \cite{reinhard2002photographic} into a common computational framework and implemented on the Graphics Processing Unit. This work realized local tone mapping by constructing a Gaussian pyramid, implement the approach in graphics hardware unit as a stand-alone HDR processing module and achieve the real-time performance.

Roch et al. \cite{roch2007interactive} propose a local tone mapping algorithm implementation \cite{ashikhmin2002tone} on graphics cards. They also present a modification of the luminance local adaptation computation, maintain the same quality appearance of the original TMO.

Zhao et al. \cite{zhao2008real} presented GPU implementations of two state-of-the-art TMOs with real-time performance. And include other six GPU-based TMOs \cite{drago2003adaptive, reinhard2002photographic, durand2002fast, ashikhmin2002tone, pattanaik2000time, pattanaik2002adaptive, colbert2007painting}, experimental evaluation is undertaken to explore which TMO is faster for hardware implementation. 

Tian et al. \cite{tiant2012gpu} proposed a real-time hardware local implementation based on global TMO \cite{duan2010tone}. They proposed a algorithm of segmenting the image into $64\times64$ independent rectangular blocks to sense local luminance. And a boundary and halo artifact elimination algorithm and a noise suppression algorithm are included in this work to improve the image quality of tone mapped image. Compare with CPU, GPU implementation can reduce the running time of a $768\times1024$ pixels image from $1.477s$ to $0.358s$ in this work.

Akil et al. \cite{akil2012real} presented a real-time GPU implementation of Irawan et al's perceptual global tonemap operator \cite{irawan2005perceptually}. The proposed system was implemented on an NVIDIA 8800 GTX GPU, they achieved real-time rendering for an HDR image dimension of $1002 \times 666$ pixels by single program, multiple data (SPMD) parallel computing.

Ure{\~n}a et al. propose a TMO and its real-time implementation in \cite{urena2013real}, this work is specially aimed as an aid system for visually impaired people who struggle to manage themselves in environments where illumination is not uniform or changes rapidly. The histogram adaptation of the luminance channel in HSV color space is used for the global tone mapping. And they propose a retina-like processing to enhance the local contrast. They achieved real-time (27 frame pre second) processing when working with $640\times480$ RGB images on NVIDIA ION2.

Eilertsen et al. presented a GPU implementation for real-time noise aware-TMO that was developed in CUDA 6.5 \cite{eilertsen2015real}. The filter design utilizes base-detail layer decomposition for tone mapping and detail enhancement is achieved through a edge-stopping non-linear diffusion approximation. On the implementation side, this TMO pipeline can process a $1980 \times 1080$ image with in 21.5 msec on a Nvidia GeForce GTX 980. 

Khan et al. \cite{khan2017tone} present a tone mapping algorithm that uses histogram of luminance to construct a lookup table for tone mapping. This kind of global tone mapping is an improvement of histogram adaptation \cite{larson1997visibility}.
Compare with CPU, GPU implementation on NVIDIA GeForce Titan Black improve the speed from 0.09 FPS to 24 FPS for images of size $2048\times1536$. The TMQI \cite{yeganeh2012objective} average of this global TMO is 0.9236 in their database.

Tsai et al. presented a GPU accelerated image enhancement method \cite{tsai2019real}, this is an improved method of their previous work \cite{tsai2011novel}. Their CUDA parallel programmed implementation can process $4096 \times 4096$ images at 64.9$\mu$s.

\section{Field Programmable Gate Array} \label{FPGA}

\begin{table*}\tiny
	\renewcommand{\arraystretch}{1.4}
	\caption{FPGA IMPLEMENTATIONS}
	\centering
	\label{table_2_2}
	\resizebox{\textwidth}{!}{
    \setlength{\tabcolsep}{0mm}{
    \begin{tabular}{|c|cccc|ccccccc|cccc|}
    \hline
    \multicolumn{1}{|c|}{} & \multicolumn{4}{c|}{\textbf{Hardware}} & \multicolumn{7}{c|}{\textbf{Cost}} & \multicolumn{4}{c|}{\textbf{Performance}} \\
    \hline
    \textbf{} & Camera & Model & \tabincell{c}{Tech-\\nology\\(nm)} & \tabincell{c}{Clock\\(MHz)} & \tabincell{c}{Latency\\(clock)} & \tabincell{c}{Power\\(mW)} & \tabincell{c}{Memory\\(bit)} & \tabincell{c}{Logic\\Elements} & DSP & Registers & Others & \tabincell{c}{Frame Size\\(pixel)} & \tabincell{c}{Speed\\(FPS)} & \tabincell{c}{Throu-\\ghput\\(Mpix/s)} & \tabincell{c}{pix/\\clock} \\
    \hline\cite{hassan2007fpga} & No &  Stratix II & 90 & 77.15 & 64 & - & 3,153,408 & 34,806 & 54 & - & - & $1024\times768$ & 60 & 47.2  & 0.61  \\
    \hline\cite{marsi2007video} & No &  Virtex II & 150 & 1.3 & - & - & 20$^\gamma$ & 17,280 & - & 890$^\delta$ & \tabincell{c}{F/F=1362$^\eta$\\Multiplier=16} & $125\times86$ & 24 & 0.258  & 0.20  \\
    \hline\cite{iakovidou2008fpga} & No &  \tabincell{c}{Stratix II\\GX} & 90 & 66.66 & \tabincell{c}{$2\times W\times H$\\$+300$} & - & 2,609,151 & 49,763$^\alpha$ & - & 43,793 & - & 2.5M  & 25    & 50    & 0.75  \\
    \hline\cite{vakili2011customized} & No &  Virtex 5 & 60 & 85 & \tabincell{c}{3,293,244\\(Total)} & - & 3,932,160 & 5,806$^\beta$ & 18 & - & - & $1024\times768$ & 25    & 19.7  & 0.23  \\
    \hline\cite{lapray2012hdr} & \tabincell{c}{EV76\\C560} &  Virtex 5 & 60 & 94.733 & \tabincell{c}{65 clock\\/line} & - & 40$^\gamma$ & 14,168$^\beta$ & 4 & 8,132 & - & $1280\times1024$ & 30    & 39.3  & 0.41  \\
    \hline\cite{kiser2012real} & No &  Spartan 6 & 45 & 75    & \tabincell{c}{12288\\@$163\mu s$} & - & 3$^\gamma$ & 874$^\beta$ & 17 & 1000$^\epsilon$ & - & $1920\times1080$ & 30    & 62.2  & 0.83  \\
    \hline\cite{urena2012real} & No &  Spartan 3 & 90 & 40.25 & - & 900 & 39$^\gamma$ & 30,086$^\beta$ & 26 & 16,545$^\epsilon$ & \tabincell{c}{BufgMux=7\\DCM=2} & $640\times480$ & 60    & 18.4  & 0.46  \\
    \hline\cite{mann2012realtime} & EyeTap &  Spartan 6 & 45 & 78.125 & - & - & 100$^\gamma$ & - & - & - & - & - & 30    & - &  \\
    \hline\cite{lapray20131} & \tabincell{c}{EV76\\C560} &  Virtex 6 & 40 & 125 & \tabincell{c}{136 clocks\\@1.2us} & 6 & 17$^\gamma$ & 16,880$^\beta$ & - & 20,192 & - & $1280\times1024$ & 60    & 78.6  & 0.63  \\
    \hline
    \multirow{2}{*}{\cite{ofili2013hardware}} & \multirow{2}{*}{No} &  Stratix II & 90 & 114.9 & nCols+12+7 & - & 68,046 & 8,546$^\alpha$ & 60 & 10,442 & - & \multirow{2}{*}{$1024\times768$} & \multirow{2}{*}{126} & \multirow{2}{*}{99} & \multirow{2}{*}{0.86 } \\
    \cline{3-12} &  &  Cyclone III & 65 & 116.5 & nCols+12+7 & 250 & 86,046 & 12,154$^\alpha$ & 36 & 10,518 & - &  &  &  &  \\
    \hline\cite{vytla2013real} & No &  Stratix II & 90 & 114.18 & - & - & 307,200 & 9,019 & 88 & 6,586 & - & 1M    & 100   & 100   & 0.88  \\
    \hline\cite{canada2013embedded} & No &  Spartan 3 & 90 & 40 & - & 897 & 720,000 & 29,007$^\beta$ & 26 & 11594 & Slice=16545 & 1M    & 100   & 100   & 0.88  \\
    \hline\cite{popovic2014performance} & No &  Virtex 5 & 65 & 214.27 & - & - & 8$^\gamma$ & 4,536$^\beta$ & 30 & 5,036 & - & $1920\times1080$ & 60    & 124.4 & 0.58  \\
    \hline\cite{shiau2014low} & No & Cyclone IV & 60 & 55.55 & 2Rows+18 & - & - & 1,784 & - & - & \tabincell{c}{Linebuff=4\\F/F=510$^\eta$} & $2560\times2048$ & - & - & - \\
    \hline
    \multirow{2}{*}{\cite{li2015low}} & \multirow{2}{*}{No} &  Cyclone II & 90 & 170.24 & - & - & 98,594 & 838 & 10 & 680 & CF=558$^\zeta$ & $1920\times1080$ & 80    & 165.9 & 0.97  \\
    \cline{3-16} &  &  Stratix III & 65 & 288.77 & - & - & 93,982 & 80 & 10 & 1,153 & CF=418$^\zeta$ & $1920\times1080$ & 140 & 290.3 & 1.01  \\
    \hline\cite{lapray2016hdr} & \tabincell{c}{EV76\\C560} &  Virtex 5 & 65 & 114 & \tabincell{c}{136 clocks\\@1.2us}  & - & 30$^\gamma$ & 6730$^\beta$ & - & - & F/F=6,378$^\eta$ & $1280\times1024$ & 60    & 78.6  & 0.69  \\
    \hline\cite{ambalathankandy2019fpga} & No &  Cyclone III & 65 & 100 & 83 & 674.25 & 87,176 & 93,989 & 28 & 26,004 & CF=67,985$^\zeta$ & $1024\times768$ & 126   & 99    & 0.99  \\
    \hline\cite{shahnovich2016hardware} & No &  Cyclone III & 65 & 100 & 2088 & 149.5 & 31,000 & 4,20 & - & 3,031 & - & $1024\times768$ & 126   & 99    & 0.99  \\
    \hline\cite{liu2016study} & No &  Stratix II & 90 & 114 & - & - & 614,440 & 9,019 & - & - & - & $1280\times720$ & 123   & 113.3 & 0.99  \\
    \hline\cite{popovic2016multi} & $-^\theta$ &  Virtex 5$\times$8 & 65 & 125 & - & 31.72W & 3 $^\gamma$ & 4,764$^\beta$ & 54 & 2,489 & - & $1024\times256\times8$ & 25    & 52.4  & 0.42  \\
    \hline\cite{nosko2017true} & \tabincell{c}{Python\\2000} & \tabincell{c}{Zynq\\XC7Z020} & 28 & 200 & - & - & 2,160,000 & 29,850 & - & - & - & $1920\times1080$ & 96    & 199   & 1.00  \\
    \hline\cite{zemvcik2017real} & \tabincell{c}{Flare\\2KSDI} & \tabincell{c}{Zynq\\XC7Z045} & 28 & - & - & 12W & 47$^\gamma$ & 24,700$^\beta$ & - & 30,405 & - & $1920\times1080$ & 30 & 62 & -  \\
    \hline\cite{popadic2017method} & Yes & - & - & 400 & - & - & - & - & - & - & - & $2M$ & 33 & 66 & 0.17  \\
    \hline\cite{nosko2018color} & \tabincell{c}{Python\\2000} &  \tabincell{c}{Zynq\\XC7Z020} & 28 & 200 & - & 8W$^\kappa$ & 34$^\gamma$ & 14,706$^\beta$ & 38 & 20,316 & - & $1920\times1080$ & 96    & 199   & 1.00  \\
    \hline\cite{yang2018local} & No &  Cyclone III & 65 & 100 & - & - & 77,408 & 13,216 & - & - & - & $1024\times768$ & 126   & 99    & 0.99  \\
    \hline\cite{yang2019mantissa} & Sensor &  Cyclone III & 65 & 100 & - & - & 107,408 & 15,471 & - & - & - & $1024\times768$ & 126   & 99    & 0.99  \\
    \hline\cite{liu2019high} & No &  Virtex 7 & 28 & 150   & - & 819 & - & - & - & - & - & $1024\times768$ & 189   & 148.6 & 0.99  \\
    \hline
    \multirow{2}{*}{\cite{ambalathankandy2019adaptive}} & \multirow{2}{*}{CX590} &  Kintex-7 & 28 & 162 & - & 453 & 9,738,000 & 9,799$^\beta$ & 21 & 15,345 & - & $1920\times1080$ & 60 & 124 & 0.77  \\
    \cline{3-16} &  & Virtex 5 & 65 & 148 & - & 804 & 7,488,000 & 10,903$^\beta$ & 22 & 12,794 & - & $1920\times1080$ & 48 & 100 & 0.84  \\
    \hline\cite{park2019low} & No &  Zynq 7000 & 28 & 148.5 & @0.241ms & - & 2,052,000 & 14,888$^\beta$ & - & 21,627 & - & $1920\times1080$ & 60 & 124 & 0.84  \\
    \hline\end{tabular}
    }}
    \tiny
    \\$^\alpha$Adaptive lookup table(ALUT) $^\beta$Lookup table(LUT) $^\gamma$Block RAM(BRAM) $^\delta$Logic cell $^\epsilon$Slice
    \\$^\zeta$Combinatorial function(CF) $^\eta$Flip-flop(F/F) $^\theta$Cell-phone cameras ($\times16$) $^\kappa$Include camera
\end{table*}

FPGAs have been a popular platform for accelerating \cite{draper2003accelerated} and prototyping many image processing pipelines that include image sensors \cite{leeser2004smart}\cite{mosqueron2007high}. FPGAs are inherently parallel, and re-programmable which makes them an ideal choice for prototyping new algorithms. For pipelined designs, it is easy to model separate hardware for different functions on an FPGA. The large resources of logic gates, RAM blocks and very fast I/O rates and bidirectional data buses altogether make them an ideal choice for prototyping the full image processing pipeline including sensor interface. Usually, the initial algorithm is designed and simulated for functional verification in software (MATLAB) and then ported to FPGA. But, simply porting these algorithms directly to a hardware platform can lead to inefficient or failure of the implementation. It is necessary to redesign and optimize the algorithm for hardware implementation keeping in mind the underlying platform. In this section we will discuss, tone mapping algorithms and novel hardware architectures that have been implemented on FPGAs, which is listed in table \ref{table_2_2}.

In 2007 Hassan and Carletta reported a grayscale HDR tone mapping system implemented on an Altera Stratix II FPGA \cite{hassan2007fpga}. The proposed solution operates on a gray scale (luminance) pixel ($P$) which is obtained as $P = 0.27R + 0.67G + 0.06B$. Their algorithm is based on Reinhard \cite{reinhard2002photographic} and Fattal's \cite{fattal2002gradient} local operators using approximation of the Gaussian pyramids. Their hardware implementation achieves 60 FPS for a $1024 \times 768$ image for pixels with 28-bit depth. They reported their output image quality in terms of PSNR, and measured an average PSNR of 34.94 dB.   

Durand and Dorsey in 2002 proposed a frequency domain based technique to tonemap HDR images \cite{durand2002fast}. This approach is similar to an earlier frequency domian filtering \cite{oppenheim1968nonlinear} in which low frequencies are attenuated more than higher frequencies. From \cite{reinhard2010high} we can understand that in an HDR image base layer tends to be low frequency and HDR, whereas the detail layer is high frequency and LDR. In the bilateral filtering \cite{durand2002fast}, a filtered output image is obtained by combining a compressed base layer with its detail layer. This approach has also been implemented on hardware. Marsi et al. \cite{marsi2007video}, used a low-pass filter to split the input image in to a low frequency base layer and high frequency detail layer. They targeted an automotive driving assistance application using a Xilinx Virtex-II FPGA, and this system included temporal smoothing to prevent flickering and color shifting. They used a $\frac{1}{4}$ down-sampled previous frame, for temporal smoothing in order to reduce memory usage and the image was stored on FPGA. Their operator achieves 24 FPS for $125 \times 86$ resolution.

Iakovidou et al., in proposed an Altera Stratix II GX FPGA is used for contrast enhancement on the images as they are received from the camera. Their algorithm is motivated by the attributes of the shunting center-surround cells of the human visual system. With a frame rate of 25 FPS, the FPGA calculates a histogram of each frame’s brightness level and transforms the image to have a stretched histogram at the output. Latency of their image processing pipeline is $2 \times W \times H + 300$ clocks, and the output image size can be up to 2.5 million pixels \cite{iakovidou2008fpga}. 

Lapray et al. \cite{lapray2011smart, lapray2012hdr, lapray20131, lapray2016hdr} in a series of publications  presented several full imaging systems using a Virtex-5 FPGA-based processing core. Their HDR imaging pipeline uses a HDR monochrome image sensor to provide a 10-bit data output and making use of Debevec and Malik's \cite{debevec2008recovering} fusion method to produces HDR video from multiple images. Using a special memory management unit \cite{lapray20131}, they can generate HDR images at the same frame rate as their camera output, which requires current frame and two previously captured frames. For HDR tone mapping they used global tonemap algorithms of Duean et al. \cite{duan2010tone} and Reinhard et al. \cite{reinhard2002photographic}. Their FPGA accelerated tonemap system achieves 60 FPS at resolution of $1280 \times 1024$ with a PSNR of 23.5 dB and a similarity index \cite{wang2004image} of 0.93.

Kiser et al. \cite{kiser2012real} mainly proposes two improvements of real-time video tone mapping system. A pre-clamping operator which the light compensation algorithm based on reference white and reference black is used to adjust the brightness of underexposure and overexposed area. They present that the pre-clamped image effectively uses more of the available output dynamic range than no-pre-clamped image. Another improvement in this work is that the tone mapping parameter curve over time is smoothed for video flicker removal. They implemented the real-time 1080p tone mapped video system on Xilinx Spartan-6 LX150T with 12288 clock latency.

Mann et al. developed and prototyped a full HDR system using a Xilinx Spartan-6 FPGA for industrial application (arc welding). Their system receives images from two head-mounted cameras, which are stored in an external memory \cite{mann2012realtime}.  Using a set of pre-stored LUT values HDR scene is constructed from three LDR images of varying exposures. They calculate the radiance with a Comparametric Camera Response Function (CCRF). For final LDR display, an estimate of photoquantity is computed from the pixel values of sequential images, which is then combined with empirical values for adjusting brightness and contrast.

Ofili et al. \cite{ofili2013hardware} proposed a hardware implementation of an exponential tone mapping algorithm. The algorithm is based on global and local compression with automatic parameter selection. It uses the global information from a previous frame and $3 \times 3$ convolution kernel to determine the local information. The output image quality has been objectively assessed using PSNR (Average = 54.27) and SSIM (0.9997) metric. Their algorithm was synthesized for both Altera Stratix II and Cyclone III FPGAs. However, this implementation is prone to halo artifacts \cite{hore2014new}.

Vytla et al. \cite{vytla2013real} developed hardware implementation of gradient domain HDR tone mapping using the Poisson equation solver inspired by Fattal's operator \cite{fattal2002gradient}. This gradient domain compression algorithm is computationally very expensive, as it seeks to solve Poisson equation. The authors of \cite{hassan2009exploiting} developed a local Fattal's operator to solve the Poisson equation locally and repeatedly, thus making it parallelize-able there by executable in real-time. The modified Poisson solver uses only local information from pixel and its $3 \times 3$ window neighbors, for computing a tone mapped pixel independent of Fattal's operator on other pixel locations with in the window. An Altera Stratix II FPGA was used to implement this algorithm, and it outputs grayscale tone mapped images with an average PSNR of 73.96 dB.

The contrast enhancement technique proposed in \cite{canada2013embedded} has targeted applications similar to \cite{urena2012real} i.e., for people with poor vision. The algorithm operates in HSI color space and is based on histogram equalization. Here, an input image is divided into 35 blocks of 5 rows $\times$ 7 columns of size 100 $\times$ 100 pixels. The histogram is computed for 64 bins. The design is implemented on multiple platforms Spartan 3, Spartan 6 and Virtex 6. They report an operating frequency of 40 MHz (Spartan 3) to 69 MHz (Virtex 6). The FPGA implementation speed-up is 15 and 7.5 times compared to CPU and GPU implementations respectively.

Popovic et al. \cite{popovic2014performance} used global TMO similar to Drago \cite{drago2003adaptive} operator. Drago's operator uses a logarithmic mapping function, to calculate displayed luminance from the ratio of world luminance and its maximum. Logarithm calculations are known to be computationally expensive, so Popovic et al., used Taylor and Chebyshev polynomials to approximate logarithms \cite{meyer2007digital}. Further, they designed a camera ring consisting of 18 independent cameras with different exposures to create panoramic HDR video \cite{popovic2016multi}. 

Li et al. \cite{li2015low} presented a FPGA hardware implementation of a contrast-preserving image dynamic range compression algorithm. The FPGA implementation is a hardware-friendly approximation to the existing FDRCLCP algorithm \cite{tsai2012fast}, and used a line buffer instead of a frame buffer to process whole image data. These advantages significantly improved the throughput performance and reduced memory requirement of the system. This implementation required only a few hardware cost and achieved high performance (Fig. \ref{fig:throughput2}).

Ambalathankandy et al. \cite{ambalathankandy2019fpga} proposed a real-time hardware platform based on hybrid TMO \cite{hore2014statistical}. This method uses local and global information from a previous frame to improve the overall contrast in the output image. Local operators are known to be prone to halo artifacts, to suppress halo in their tone mapped images they implemented a halo-reducing filter \cite{hore2014new}. The proposed system was targeted for an Altera Cyclone III FPGA, processing images in the luminance channel and producing output images with an average PSNR = 57.27dB and SSIM = 0.9969.

In late 90's mantissa-exponent representation were chosen to record digitized samples from the image sensor. Multiple sampling for exponentially increasing exposure times $T, 2T, 4T,... 2^{k}T$ were coded with m bits. Each pixel sample were represented with m+k bits, where the exponents ranged between 0 to k and mantissa was m-bits. This representation extends the dynamic range by $2^{k}$ times, and provides m-bits resolution for each exponent range of illumination while incurring lower memory costs \cite{david1999640, spivak2009wide}. Shahnovich et al. used this representation in their FPGA implementation \cite{shahnovich2016hardware}. Their TMO makes use of a simple logarithmic compression module for HDR images using this mantissa-exponent representation, they treat every input pixel to be 24 bit wide, where 16 bits is used for the mantissa and 8 bits for the exponent representation. Yang et al. \cite{yang2019mantissa} also considered the mantissa-exponent pixel to obtain a refined histogram which utilizes density and dispersion information to construct a piece-wise model for their tone mapping function. This FPGA implemented TMO system can process $1024 \times 768$ images with 10 bits mantissa and 3-bits exponent.

Nosko et al. \cite{nosko2017true, nosko2018color} described a fast implementation of HDR merge processing by multiple exposure and a local TMO involving bilateral filtering \cite{durand2002fast}. This work also propose an application of de-ghosting method, which is dedicated for FPGA implementation. Compared with the use of Gaussian filter to detect local luminance, bilateral filter can preserves sharp edges, but also require more hardware cost. This work can output one pixel per clock on Zynq 7020 FPGA at 200 MHz.

Zemvcik et al. \cite{zemvcik2017real} presented a real-time HDR video acquisition and compression system built using FPGA that include captures multiple exposure HDR video \cite{myszkowski2008high}, merges the standard range frames into the HDR frames and compresses the HDR video using TMO \cite{banterle2008hdr}.

Popadic et al. \cite{popadic2017method} proposed an exposure fusion (Fig. \ref{fig:EF1}) hardware implementation. This work implemented on FPGA inserted into standard industrial digital camera. The tone mapped images were obtained by using weighted summation of three images captured with different exposure times. The weight coefficient of each image was calculated according to the visibility threshold curve \cite{chou1995perceptually}. This work performs calculations on the global image level, and implements real-time (30ms/image) processing with 2M pixel images at 400 MHz.

Yang et al. implemented a segmentation-like TMO, here the input HDR image is divided into m$ \times $n blocks \cite{yang2018local}. The algorithm uses block-interpolated minimum and maximum values to determine the compression values. The Cyclone III FPGA implementation operates in a pixel-by-pixel fashion in the logarithmic domain processing 1024$\times$768 images operating at 100 MHz.   

Liu et al. reported a retina-inspired \cite{zhang2016retina} tone mapping algorithm by implementing horizontal and bipolar cell stages on a Xilinx Virtex 7 FPGA \cite{liu2019high}. The design operated at 150 MHz consumed 819 mW power while processing a $1024 \times 768$ image which corresponds to an energy efficiency of 544453 pixels/mW/s.

Ambalathankandy et al. presented a global and locally adaptive tone mapping algorithm and its FPGA implementation on a Xilinx Kintex 7 to process Full HD images \cite{ambalathankandy2019adaptive}. They make use of a non-traditional white color space in their algorithm, a RGB color image can be transformed to this space as $W = \sqrt{\frac{R^{2}+G^{2}+B^{2}}{3}}$.Their tone mapping function, is based on local histogram equalization, controls global and local characteristics individually and can manage light/dark halos. By a weighted function, they demonstrate noise suppression in tone mapped images. To reduce memory access and latency, they use a downscaled previous frame (1/64).

Park et al. \cite{park2019low} presents the FPGA implementation of the efficient naturalness restoration algorithm \cite{shin2015efficient}. The proposed FPGA design used small line buffers instead of frame buffers, applied a concept of approximate computing for the complex Gaussian filter, and designed a new nontrivial exponentiation operation to minimizes hardware resources while maintaining quality and performance. The design supports a throughput of 60 frames/s for a $1920 \times 1080$ image on Zynq 7000 at 148.5MHz. In this paper, they normalized each resource of different types (LUT, register, block memory, and external memory)into memory bits   to the user guide of each FPGA fabric \cite{Virtex-4, 7Series, FPGALogic}. And compared the normalized comparison metric with other FPGA tone mapping implementations which are implemented on different models.

\section{Application Specific Integrated Circuits}

As we saw earlier, building full custom solutions on ASIC is expensive and time consuming. However, there may be applications with specific demands on power consumption, size of the hardware or some security aspects that may need specific ASIC implementations. ASIC is useful for building dedicated hardware which can be integrated on a single chip with sensor. They can clock at very high speed, and be faster than a commonly used camera. There by, making it possible to process the data and be sent out for further actionable controls. In table \ref{table_2_3} we have listed TMO algorithms realized on ASIC platform, and details of which follows next. 

\begin{table*}\scriptsize
  \renewcommand{\arraystretch}{1.4}
  \caption{ASIC IMPLEMENTATIONS}
  \centering
  \label{table_2_3}%
	\resizebox{\textwidth}{!}{
    \setlength{\tabcolsep}{0mm}{
    \begin{tabular}{|c|ccccc|ccc|cccc|}
    \hline\multicolumn{1}{|c|}{} & \multicolumn{5}{c|}{\textbf{Hardware}} & \multicolumn{3}{c|}{\textbf{Cost}} & \multicolumn{4}{c|}{\textbf{Performance}} \\
    \hline\textbf{} & CMOS Sensor & Foundry & \tabincell{c}{Technology\\($\mu m$)} & Monolithic & \tabincell{c}{Clock\\(MHz)} & \tabincell{c}{Area(mm$^2$)} & \tabincell{c}{Power\\(mW)} & \tabincell{c}{Gate\\Counts} & \tabincell{c}{Frame Size\\(pixel)} & \tabincell{c}{Speed\\(FPS)} & \tabincell{c}{Throughput\\(Mpix/s)} & \tabincell{c}{pix/\\clock} \\

    \hline\cite{chiu2011real}& No    & TSMC & 0.13 & No & 100   & \tabincell{c}{$2.85\times2.85$(Core)/\\$3.74\times3.74$(bond)} & 177.1478 & 769,620 & $1024\times768$ & 60 & 47.2 & 0.47 \\
    \hline\cite{punchihewa2011review}& Yes & - & - & No & - & $1.98\times3.88$(Chip) & - & - & $64\times64$ & - & - & - \\
    \hline\cite{sicard2013cmos}& No & AMS & 0.35 & No & - & - & - & - & $256\times256$ & - & - & - \\
    \hline\cite{shiau2014low}& No    & TSMC & 0.13 & No & 200   & $1.835\times1.835$ (Core) & -   & 15,742 & $2560\times2048$ & 37    & 194 & 0.97 \\
    \hline\cite{vargas2014151}& Yes (151dB) & AMS & 0.35 & Yes & - & $7.33\times6.78$(Core) & 111.2 & - & $180\times148$ & 1205  & 32.1 & - \\
    \hline\cite{gouveia2014reconfigurable}& Yes (120dB) & Lfoundry & 0.15 & in-pixel & - & - & - & - & - & - & - & - \\
    \hline\cite{mughal2015fixed}& No & AMS & 0.35 & in-pixel & - & - & - & - & - & - & - & - \\
    \hline\cite{fernandez2015single}& Yes (102dB) & - & - & No & - & - & - & - & - & - & - & - \\
    \hline\cite{shi2016tone}& Yes & - & 0.15 & in-pixel & 52 & -   & 54.72 & - & $1024\times768$ & 66    & 51.9 & 1 \\
    \hline\cite{chen2016analog}& No    & TSMC & 0.35 & Yes & - & 0.039 & 41 & - & - & - & - & - \\
    \hline\cite{guicquero2016algorithm}& Yes (CIS) & - & - & Yes & - & - & - & - & $512\times512$ & - & - & - \\
    \hline
    \end{tabular}%
    }}
\end{table*}%

A $64 \times 64$ pixels image sensor using 5 transistors APS per pixel with adaptive integration method is proposed by Punchihewa et al. \cite{punchihewa2011review}. Their They developed a tone compression algorithm that is based on improved local histograms and using differential-luminance histograms. The proposed system was tested using a 16 bit image raw data.

Sicard et al. \cite{sicard2013cmos} described an analog model of the Gamma correction method proposed by Meylan et al. \cite{meylan2007model} for local tone mapping. Sicard's method improves upon digital normalization of the pixel output and this is in line with the Michaelis Menten law, however this study reports few preliminary results only.

Shiau et al. developed a transformation domain-based method to estimates illumination by a bi-dimensional empirical mode decomposition \cite{shiau2014low}. They adjusted the image contrast by gamma correction there by avoiding over-enhancement in the output image. The algorithm was implemented on multiple hardware platforms, SYNOPSYS DV was used to synthesize the design with TSMC 0.13$\mu$m technology. The synthesized design core size was 3.367$mm^{2}$ with a gate count of 11.6K and with a clock period of 5 ns and can achieve a processing rate of 200 Mpixels/s. The design implemented on FPGA consumed 1,784 logic elements and operated with a clock at 55 MHz. 

Vargas-Sierra et al. \cite{vargas2014151} developed a proof-of-concept HDR CMOS image sensor that implemented a global tonemap compression during image capture operation. The system has been conceived as a complete Vision System-on-Chip (VSoC) with a core area of $7.33mm \times 6.78mm$ was fabricated on $0.35 \mu m$ opto-flavored technology. This system achieved video rates for QCIF resolution images with 25 bits/pixel and tonemap them to 7 bit/pixel for display. But, an off-chip processing is required to compute image histogram before the final display \cite{lenero2017wide}.

Gouevia et al. in \cite{gouveia2014reconfigurable} present a new programmable pixel based on variable integration time. In this proof-of-concept, the integration time is a function of the light intensity and is an user controllable signal. Using this programmable pixel, they try to simulate the monotonic non-linear response of tonemap functions. They studied three different operators \cite{reinhard2002photographic,reinhard2010high,drago2003adaptive} and tested the implementation output for various test images \cite{vcadik2008evaluation,kuhna2011method}.

Mughal et al. in \cite{mughal2015fixed} reported a new pixel with inbuilt TMO, this operator was based on, the Reinhard's photographic TMO \cite{reinhard2002photographic}. The performance of pixel circuits are limited by the fixed pattern noise(FPN) which is mainly due to the variations between the responses of individual pixels  within an array of pixels. Mughal et al. devise a calibration technique to to correct the fixed pattern noise in pixels which can produce a tone mapped response.

Fernandez et al. \cite{fernandez2015single} designed a vision sensor with dual photodiode pixel, of these two photodiodes large one enabled them to capture high sensitivity and the small low sensitivity in the same exposure. Also, the large photodiode sensed the pixel value and the small photodiode achieved a tunable balance between local and global adaptation. By taking inspiration from earlier work \cite{vargas2014151}, a global operator for HDR tone-mapping compression based on an online evaluation of the image histogram is implemented in OpenCV. The proposed system fabricated in 0.18$\mu$m CMOS technology can capture an image with dynamic range up to 102 dB. 

Shi et al. \cite{shi2016tone} modeled Reinhard's \cite{reinhard2002photographic} photographic tone reproduction operator in analog domain using Verilog-A. Their  global tonemap operator achieves 60fps for $1024 \times 768$ monochrome images while consuming 54.27 mW power. They also demonstarted the usefulness of this work to  improve the dynamic range of CT images that are affected by overexposure artifacts \cite{shi2016analog}.

Chen et al. \cite{chen2016analog} designed an analog current-mode joint global and local tonemap operator based on \cite{ofili2012depth}. From simulations they report, their TSMC CMOSP35 implementation consumed 41 mW power from a 3.3V supply and it requires 1$\mu s$ processing time per pixel.

Compressive sensing (CS) based HDR imaging was proposed by Guicquero et al. \cite{guicquero2016algorithm}. Their method utilizes cellular automaton for a scalable and low-complexity column based compressive sensing. They describe an imager architecture for $512 \times 512$ pixel and a dedicated global tonemap reconstruction algorithm. The algorithm iteratively evaluates image contrast. To emphasize low-light details and to suppress noise the iterative tonemap operation flattens out the histogram.

\section{Customized Processors and Other Hardware Platforms} 

Application Specific Instruction set Processors (ASIP) has lately emerged as an attractive platform to implement signal processing algorithms \cite{amano2006survey}. ASIPs are highly customizable, and embedded with domain-specific hardware accelerators, and these hardware accelerators are strongly coupled with the processor pipeline and are easily accessed by custom instructions. An ASIP can be systematically customized for implementing a specific application due to the availability of custom instructions, availability of optimized domain-specific blocks, and other parameterizable options \cite{karuri2011application}. ASIPs have been used for real-time Retinex-like image and video enhancement applications \cite{saponara2007application}\cite{saponara2007algorithmic} and global tone mapping \cite{vakili2011customized}. 

Saponara et al., in \cite{saponara2007application} presented a programmable class of Retinex-like filters, based on the separation of the illumination and reflectance components. The dynamic range of an input image is modified by applying a non linear function to compress the illumination layer and enhancing the details in the reflectance layer. They proposed 42 new ASIP instruction set which perform: color conversion, nonlinear transformations, arithmetic computations, memory accesses, initialization, and loop control. The final design was synthesized using Synopsys in $0.18\mu m$ 1.8V CMOS standard cells technology. Their system could process a $256 \times 256$ video at up to 29 fps.

Vakili et al., in \cite{vakili2011customized} proposed an implementation of Reinhard's global tone mapping algorithm \cite{reinhard2002photographic} on a customized LTRISC, which is a 32-bit RISC-like processor model provided with Synopsys Processor Designer. Using the LISA ADL, they developed three instructions to based on a specific low cost technique presented in \cite{mitchell1962computer} to calculate (i) luminance, (ii) logarithm and (iii) maximum luminance. They manually determined word length for all the intermediate variables. The system was synthesized for a Xilinx Virtex-5 FPGA and achieves 25 FPS for image size of $256 \times 192$ pixels using a clock frequency of 85 MHz. The system outputs image with an average PSNR of 50dB. 

ASIPs have comparatively better energy and area efficiencies than the digital signal processors (DSP) and general purpose processors (GPP) \cite{lee2009algorithm}, while it does offer the flexibility and programmability for algorithm re-design/upgrades and bug-fixes. However, this programmability feature comes at a price of increased area (loop control, registers, etc.), when compared to ASICs. Development of an ASIP system is laborious, as is evident from the 2 man-months effort spent for Architecture Description Language (ADL) design in \cite{saponara2007application}, further the system development not only involves design and verification of ASIP architecture but also the construction of the associated software tools such as assembler, compiler, debugger and instruction set simulator \cite{karuri2011application}.

Chiu et al. \cite{chiu2011real} developed a tone-mapping processor based on an ARM core with an application-specific integrated circuit (ASIC). Their processor includes a modified global photographic tone mapping and a block-based gradient domain compression, based on algorithms proposed by Reinhard et al. \cite{reinhard2002photographic} and Fattal et al. \cite{fattal2002gradient}, respectively. The processor can run at a 100 MHz clock rate and can compress $1024 \times 768$ HDR images at 60 fps. However, this approach does not offer the flexibility like others \cite{saponara2007application}\cite{saponara2007algorithmic}\cite{vakili2011customized} since some critical modules are implemented on an ASIC core that occupies $8.1mm^{2}$ of physical area in $0.13\mu m$ TSMC technology.

Ure\~na et  al. \cite{urena2012real} implemented their own optimized tone mapping algorithm. Their algorithm operates in HSV color space and performs histogram equalization of the brightness (V) channel. They also perform local detail enhancement by a $7 \times 7$ window convolution. The output brightness is a simple combination of the convolution output and histogram equalization. Their algorithm also included a glare mitigation filter, as they intended to develop this system for a low power battery operated device for visually impaired people. They prototyped it on two different platforms, one on a Xilinx Spartan III FPGA and a Nvidia ION2 GPU. Their proposed implementations delivered real-time performances of 30 FPS and 60 FPS on GPU and FPGA respectively for $640 \times 480$ image resolution with an average PSNR measure of 17.35 dB. In their subsequent work they implemented a filter to attenuate glares in images, because people with poor vision have difficulties in adapting to illumination changes \cite{urena2013real}. 

Narashima and Batur \cite{narasimha2015real} presented a real-time tone mapping algorithm implemented in software using Texas Instrument Davinci media processor \cite{TID2020}. In their implementation, a luminance image is divided into overlapping blocks and a block mean pixel value is computed for each of the blocks. Several local tonemap functions are used to adapt each of the sub-blocks. A limitation of this algorithm is the choice of the block size and amount of overlap which may have to  be determined empirically, there by limiting its practical application as it can significantly affect its quality and performance.

\chapter{Quantitative Comparison}

\section{Hardware}

\begin{figure*}[]
\centerline{\includegraphics[width=\textwidth, trim=110 10 -10 40, clip]{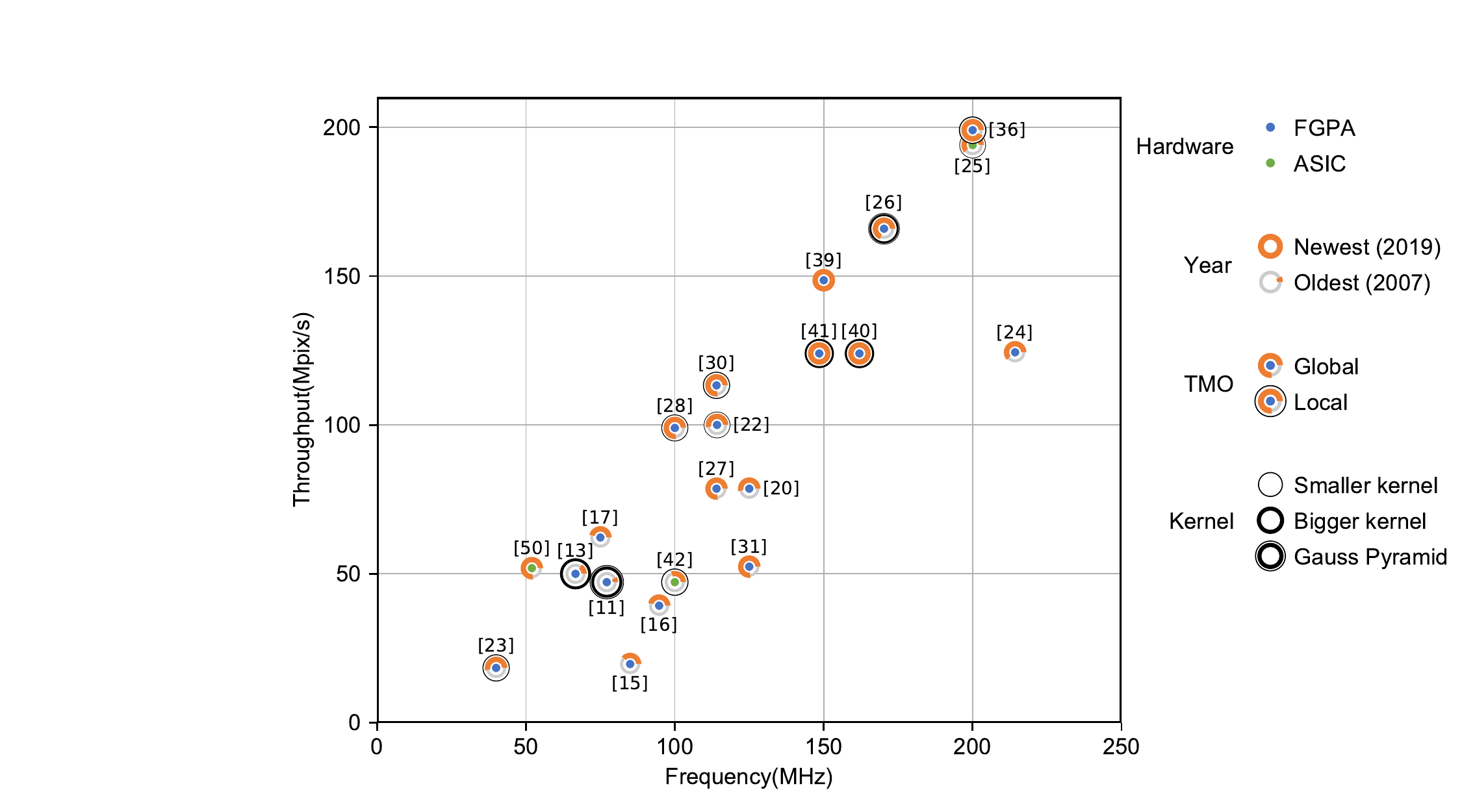}}
\caption{TMOs Performance Measurement: Throughput versus operating frequency is an important measure for real-time performance.}
\label{fig:throughput1}
\end{figure*}

\begin{figure*}[]
\centerline{\includegraphics[width=\textwidth, trim=110 10 -10 40, clip]{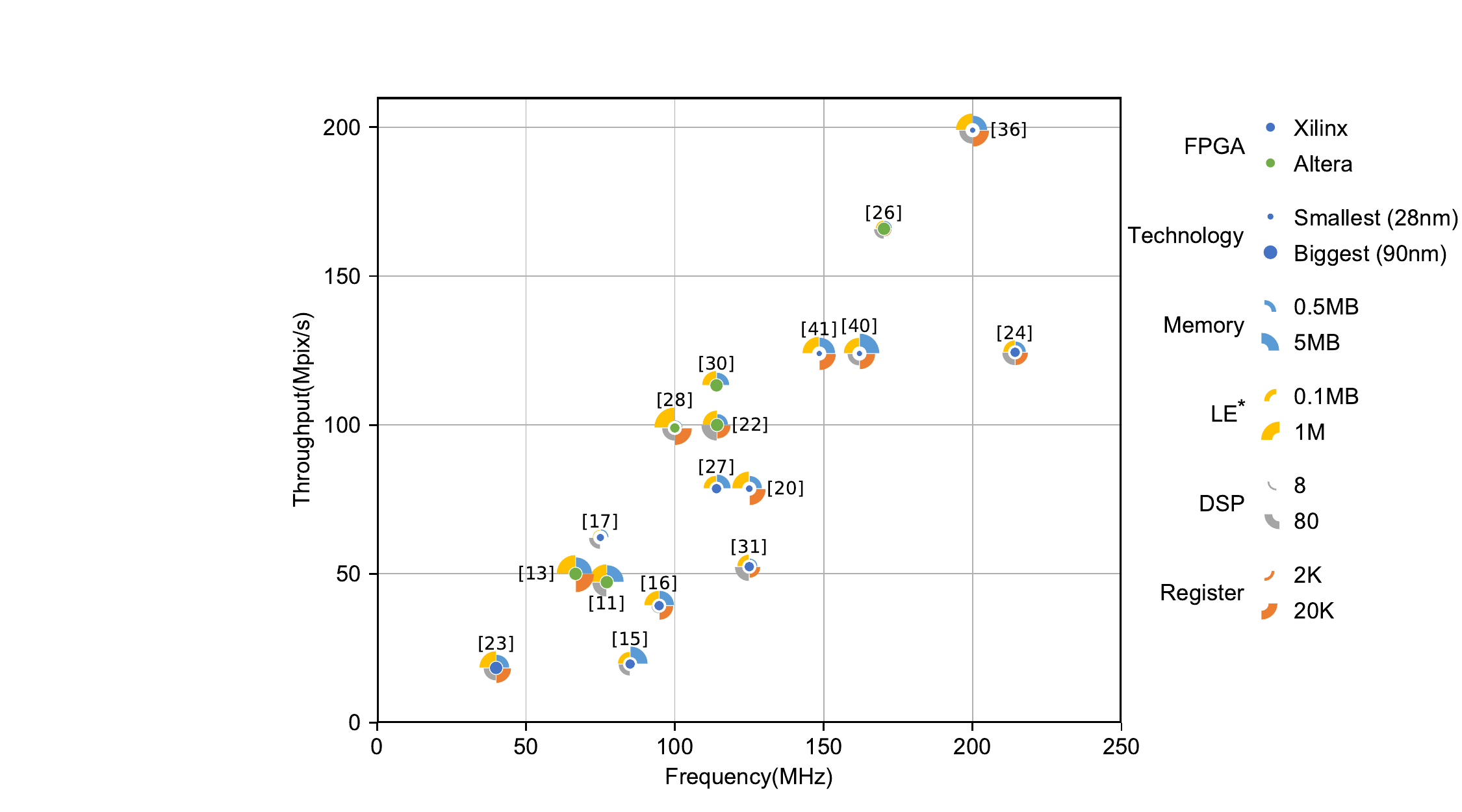}}
\caption{Throughput versus Hardware Cost: TMO cost is evaluated in terms of memory, DSP, logic elements and registers. Global TMOs are usually light-weight in comparison to local TMOs.}
\label{fig:throughput2}
\end{figure*}

\begin{figure}[!t]
\centerline{\includegraphics[width=8cm, trim=15 0 40 30, clip]{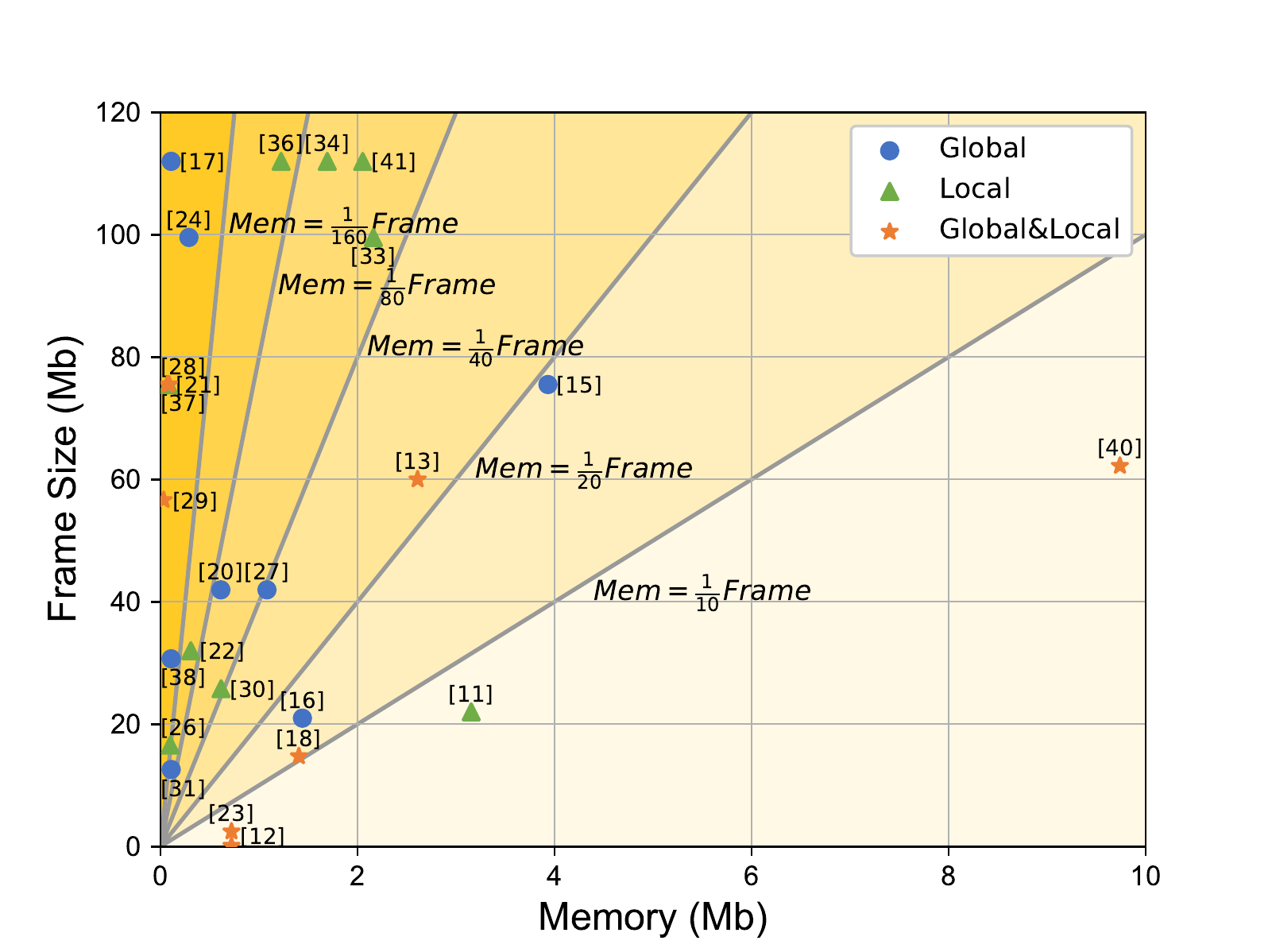}}
\caption{Memory Cost: Plot illustrates memory cost of tone mapping algorithms versus output images. Typically global TMOs have lower memory requirements than local TMos.}
\label{fig:memory}
\end{figure}

As stated earlier, one of the main objective of our survey is to determine the quality of various TMOs. In Fig 5.13 we present ASIC and FPGA implementations grouped on a map. In the map a blue dot imply FPGA implementation, and green one is ASIC. The positions on this map highlights these implementations according to their frequency and throughput. We computed throughput as shown in Eq. \ref{eq:thoughput}

\begin{equation}
Thoughput=Height_{Image} \times Width_{Image} \times FPS
\label{eq:thoughput}
\end{equation}

Frequency and throughput are very important parameters, as it allows us to predict the algorithm's effectiveness for processing HDR images of wide-ranging resolutions. In other words, more throughput per cycle implies that the algorithm is faster, and is well optimized. The orange circle in Fig \ref{fig:throughput1} represents the year of publication. A paper which has full circle (in orange color) means chronologically newer paper, and arguably this paper would have improved design architecture than those in previous papers, thereby, realizing some design optimization. The black circle gives information of tone mapping. The width of black circle corresponds to the size of kernel that is used to calculate the local luminance. Works based on global TMO are without black circle because they do not use any local statistics. With respect to local operators, some of those works have calculated local luminance by using Gauss Pyramid, such implementations are represented as double black circles and their performances are also included in Fig \ref{fig:throughput1}. Figure \ref{fig:throughput2} illustrates the hardware cost of FPGA tone mapping implementations. Like in Fig \ref{fig:throughput1} the location of each point is grouped with respect to the system throughput and frequency, so that same position corresponds to same paper in both figures of merit. The color of center circle of each point informs about the FPGA manufacturer (ALTERA/XILINX) that were used in research. And the size corresponds to the FPGA's manufactured technology. Smaller size represent that they use newer 22nm technology.

To compare the TMO algorithms implemented using different FPGA families, we follow the hardware normalization strategy proposed by Park et al. \cite{park2019low}. For algorithms implemented on Xilinx Virtex-4, each LUT and register consumed can be substituted with 16-bit memory. For systems that are implemented on a Virtex-7/Zynq7000, and those seven series FPGA's LUT and registers are equalized with 32-bit memory. For Altera Cyclone III based systems, according to an article analyzing the difference between the two FPGA fabrics \cite{FPGALogic}, one unit of the logic element used in Cyclone III is 1.3 times larger than one unit of LUT used in Virtex-4, so the resource utilization is converted into the estimated amount of LUTs of Virtex-4. This method of normalizing hardware resources between different FPGA fabric usages for comparison has also been adopted by Choi et al. \cite{choi2018high}. The size of the circle sector area is proportional to the hardware cost. Different colors represent different type hardware. By studying Fig \ref{fig:throughput1} we can see that recent works \cite{shiau2014low,nosko2018color,ambalathankandy2019adaptive} have high throughput.With streaming applications like tone mapping which continuously process data, throughput is the most interesting design aspect as it will define the performance of the tone mapping application. More throughput means that more data can be processed in the same instant of time. To improve the performance of slower algorithms, one of the most effective ways to is by adding extra pipeline stages. FPGA designs have a synchronous nature consisting of delay elements and logic, which means that it highly benefits from extra pipeline stages. 

Another interesting design aspect is the memory cost, Fig \ref{fig:memory} shows the relationship between TMO HW implementation memory and output frame size. Frame size is computed as shown in Eq. \ref{eq:frame}. From the Fig .\ref{fig:memory} we can observe that, as expected global TMOs require less memory than local TMOs. In the Fig \ref{fig:memory}, we have grouped the algorithms based on the performance as which of these implementations are better designed to reduce memory cost. Recent local and global TMO works \cite{zemvcik2017real,nosko2018color,park2019low} report low memory usage by highly optimizing their designs. For example, Park et al., designed a frame-less TMO system, and only used a small line buffer\cite{park2019low}. They further optimized their design by building an approximate convolution block of a $29 \times 29$ Gaussian filter. A conventional 2-D filtering scheme would have required $29 \times 29$ convolutional operations between a pixel and its coefficient. They implemented it by using two 1-D separable filters operating vertically and horizontally, thereby reducing the number of operations to $29 \times 1$ plus few additional adders.    

\begin{equation}
Frame_{Size}=Height_{Image} \times Width_{Image} \times Datawidth
\label{eq:frame}
\end{equation}

\section{Data Conversion for Optimal Hardware Specification}\label{FPA}

\begin{figure}[!b]
\centerline{\includegraphics[width=7.5cm, trim=-40 0 0 0]{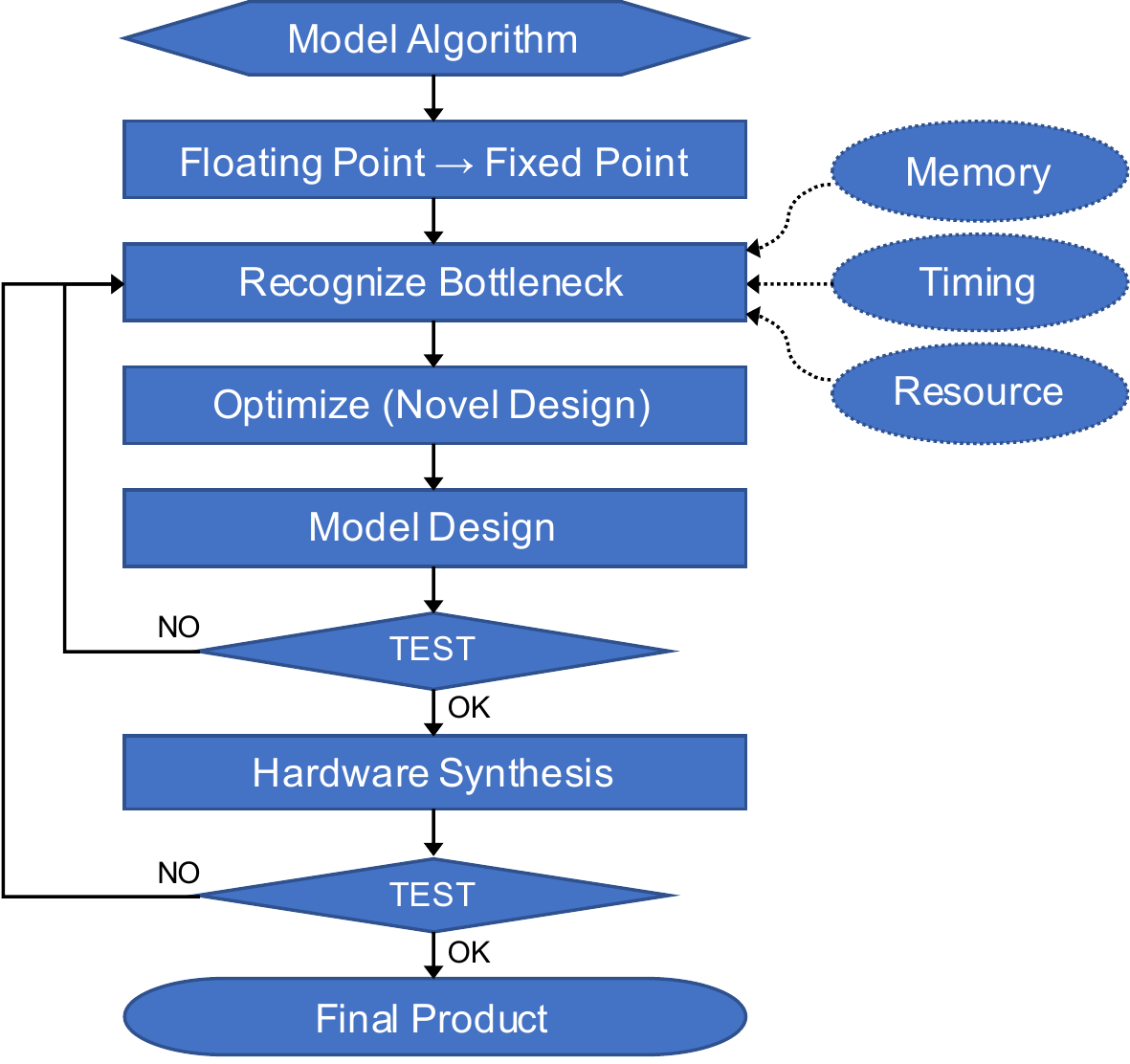}}
\caption{Flowchart listing common approach to adapt a software algorithm to a hardware platform.}
\label{fig:flowchart}
\end{figure}

The cost of modern electronic devices is usually measured in terms of silicon area (chip footprint), power consumption and algorithm/application execution time. Engineer's strive hard to keep these three factors to a minimum while attaining all the system objectives. Balancing these goals are extremely challenging in nature, and usually a delicate trade-off between system performance and cost has to be planned in advance. Therefore, it is very important to make careful decisions in every design step to ensure the best possible performance of the entire system. Direct porting of a software tone mapping algorithm to any hardware platform will be inefficient and even may lead to system failures. The various stages of the design flow of a digital signal processing application are described in Fig \ref{fig:flowchart}. As a first step the algorithm is designed, simulated and thoroughly tested using DSP software tools like MATLAB/Simulink (Mathworks Inc) or others like Scilab etc. These software implementations achieve high degree accuracy as the algorithms are described with floating-point arithmetic. However, the  choice of arithmetic operators used to implement the algorithms, has a decisive impact on the cost-performance trade-off. On a hardware platform fixed-point arithmetic operators are ideal choice as they require low area (footprint), low power, and have low latency \cite{shi2004floating}.

But, the floating-point to fixed-point conversion process is an optimization problem which derives the data word-length \cite{shi2003automated}. This adaptation has to be carefully crafted as this leads to a trade-off between cost and performance. Reducing the data-width leads to significant quality degradation and therefore appropriate assessment is required to various stages to ensure acceptable performance. Also, this conversion process is not straightforward and may require significant design effort. But it is encouraging to note that there are automatic floating-point to fixed-point conversion tool which can significantly speed-up this process \cite{Matlab}. Next step in the design cycle is to determine the data-width for integer and the fractional part. It is absolutely necessary to establish this step by performing detailed numerical analysis as this will effect the overall performance due to accuracy of the algorithm. The integer part determines the dynamic range of the data and fractional part determines the numerical accuracy. The integer part determines the minimum and maximum values and, the designer has to accurately determine the number of bits required to represent the range depending upon the application thereby avoiding any overflows. 

Recently, some High Level Synthesis tools (HLS) like Intel HLS, Cadence Stratus etc. have emerged, which can directly generate register transfer level (RTL) implementations from a C/C++ fixed-point specification of the algorithms \cite{intelHLS, Xilinx, Mentor, Cadence}. Use of these tools can speed-up the re-design effort, but the designs may not be fully optimized. Other issues that affect the direct adoption of fixed-point arithmetic from the floating-point designs using HLS:

\begin{itemize}
	\item Bit growth: When performing arithmetic operations like addition and multiplication, designer has to maintain adequate width on datapath. Over-constrained designs can lead to loss of accuracy and relaxed approach could lead to 
	waste of resources.   
	\item Complex functions: When performing various complex mathematical functions (transcendental) in floating-point algorithm (for example in MATLAB/C/C++) is precisely defined. However, accurate equivalent hardware 
	micro-architectures may not be available. 
\end{itemize}

\section{Image Quality Assessment}

\begin{table}[!b]\scriptsize
  \renewcommand{\arraystretch}{1}
  \caption{Image Quality Metrics}
  \centering
  \label{tab:metrics}
  \begin{threeparttable}
    \setlength{\tabcolsep}{1.25mm}{
    \begin{tabular}{m{0pt}rlcc}
    \hline \rule{0pt}{12pt} & \multicolumn{2}{c}{\small Image Quality Metrics} &\small Ideal Value &\small \hspace{-1mm}Reference\\
    \hline \rule{0pt}{12pt} & SSIM &\hspace{1mm} Structural Similarity Index & 1 &\hspace{-1mm}\cite{wang2004image}\\
    \rule{0pt}{12pt} & TMQI &\hspace{1mm} Tone Mapped Image Quality Index & 1 &\hspace{-1mm}\cite{yeganeh2012objective}\\
    \rule{0pt}{12pt} & PSNR &\hspace{1mm} Peak Signal to Noise Ratio & higher is better & \hspace{-1mm}$20log_{10}(\frac{MAX}{RMSE})^\ddag$\\
    \rule{0pt}{12pt} & RMSE &\hspace{1mm} Root Mean Square Error & 0 & \hspace{-1mm}$\sqrt{MSE}$\\
    \rule{0pt}{12pt} & MSE &\hspace{1mm} Mean Square Error & 0 & \hspace{-1mm}$\frac{SSE}{mn}$\\
    \rule{0pt}{12pt} & SSE &\hspace{1mm} Error Sum of Squares & 0 & \tiny \hspace{-1mm}$\sum\limits_{i=0}^{m-1}\sum\limits_{j=0}^{n-1}[I_{ij}-G_{ij}]^2$\\
    \rule{0pt}{12pt} & RMS\% &\hspace{1mm} Root Mean Square Percentage & 0 &\tiny \hspace{-1mm}$\frac{1}{mn}\sum\limits_{i=0}^{m-1}\sum\limits_{j=0}^{n-1}[\frac{I_{ij}-G_{ij}}{I_{ij}}]^2$\\
    \rule{0pt}{12pt} & R-squared &\hspace{1mm} Coefficient of Determination & 1 &\hspace{-1mm} \cite{watkins2010statistics,cameron1997r}\\
    \rule{0pt}{12pt} & DE &\hspace{1mm} Discrete Entropy & higher  is better&\hspace{-1mm} \cite{shannon1948mathematical}\\
    \rule{0pt}{12pt} & EBCM &\hspace{1mm} Edge-Based Contrast Measure & higher is better &\hspace{-1mm} \cite{beghdadi1989contrast}\\
    \rule{0pt}{12pt} & CEF &\hspace{1mm} Color Enhancement Factor & higher is better &\hspace{-1mm} \cite{mukherjee2008enhancement}\\
    \rule{0pt}{12pt} & $\Delta E_{HS}$ &\hspace{1mm} Color Similarity & 0 &\hspace{-1mm} \cite{tsai2012fast}\\
    \rule{0pt}{12pt} & UQI &\hspace{1mm} Universal Quality Index & 1 &\hspace{-1mm} \cite{wang2002universal}\\
    \rule{0pt}{12pt} & AMBE &\hspace{1mm} Absolute Mean Brightness Error & 0 &\hspace{-1mm} \cite{chen2003minimum, celik2011contextual}\\
    \rule{0pt}{12pt} & EME &\hspace{1mm} Enhancement Measure & higher is better &\hspace{-1mm} \cite{agaian2007transform, panetta2008human}\\
    \rule{0pt}{12pt} & GCC &\hspace{1mm} Global Contrast Change & higher is better &\hspace{-1mm} \cite{smith2006beyond}\\
    \rule{0pt}{12pt} & DM\% &\hspace{1mm} Detail Maintenance Percentage & higher is better &\hspace{-1mm} \cite{liu2016study}\\
    \hline
    \end{tabular}%
    }
    \begin{tablenotes}
    	\item[\ddag] $MAX$ is the maximum possible pixel value of the image.
	\end{tablenotes}
	\end{threeparttable}
\end{table}%

Image Quality Assessment (IQA) is plays vital role at many levels of the design cycle, and an early assessment is inevitable to prove the usefulness of the algorithm. subjective user study is the most reliable means to measure image quality. However, it is not always feasible for practical reasons. During the hardware development stages, objective image quality metrics are used to evaluate the system performance and analyze if there is more room for optimization. IQA studies has been actively carried out  and there are several quality metrics in literature \cite{lin2011perceptual, liu2011image, mantiuk2012comparison}. These IQA methods try to accurately predict the subjective preferences of a common human user by surveying the perceived quality of visual data presented to the user. These user studies are grouped as full-reference and no-reference IQA methods depending upon the availability of ideal reference images, with which the images (here tone mapped images) will be compared. 

For algorithms implemented on hardware, IQA is measured in terms of image distortion which is caused by the approximations due to floating-point to fixed-point translation. From our survey we observe that in literature PSNR is a preferred metric to measure the pixel value distortion between software and hardware tone mapped images. Other metrics also have been used and they are all listed in table \ref{tab:metrics}. 

As stated earlier, among them the most commonly used metrics are peak signal to noise ratio (PSNR), structural similarity (SSIM) index and tone mapped image quality index (TMQI).

PSNR is a well-known quality metric used to evaluate the image quality by the mean-square error (MSE). Similar to PSNR, some traditional metrics based on square error were also been used for image quality assessment in tone mapping hardware implementation, such as root mean square error (RMSE), mean square error (MSE), error sum of squares (SSE), root mean square percentage (RMS\%), R-squared and so on. The PSNR value approaches infinity as the MSE approaches zero, this shows that a higher PSNR value provides a higher image quality. But the PSNR perform badly in discriminating structural content in images since various types of degradations applied to the same image can yield the same value of the MSE \cite{hore2010image, wang2009mean}.

Therefore, some more elaborate methods attempt to incorporate structural information in IQA. Wang et al. \cite{wang2004image} proposed SSIM based on human visual perception for measuring the similarity between two images, and Yeganeh et al. \cite{yeganeh2012objective} proposed an objective quality assessment algorithm for TMO and named it as TMQI which is based on SSIM and naturalness. In TMO hardware implementation works, SSIM is used for pixel-to-pixel calculating the similarity between tone mapped images by software operator and hardware implementation. During hardware implementation, there have some losses due to approximate calculations, and SSIM is used to measure these losses. When SSIM is close to 1, it indicates that losses in hardware implementation is small. And TMQI is used for measure the tone mapped quality from HDR images to LDR images which were processed by hardware.

Some previous works have used some special metrics for image quality assessment. Such as, discrete entropy (DE)\cite{shannon1948mathematical} for measuring the degree of details in images, edge-based contrast measure (EBCM)\cite{beghdadi1989contrast} for evaluating the contrast of images, color enhancement factor (CEF)\cite{mukherjee2008enhancement} for measuring colorfulness, $\Delta E_{HS}$ for color similarity, universal quality index (UQI)\cite{wang2002universal} for the similarity between two images, absolute mean brightness error (AMBE)\cite{chen2003minimum, celik2011contextual} for the preservation of the original image brightness, enhancement measure\cite{agaian2007transform, panetta2008human} and global contrast change\cite{smith2006beyond} for contrast. Some works also present their image quality metrics, like detail maintenance percentage \cite{liu2016study}.

\section{Hardware Specification versus Image Quality}

\begin{figure}[!b]
\centerline{\includegraphics[width=8cm, trim=5 0 0 0, clip]{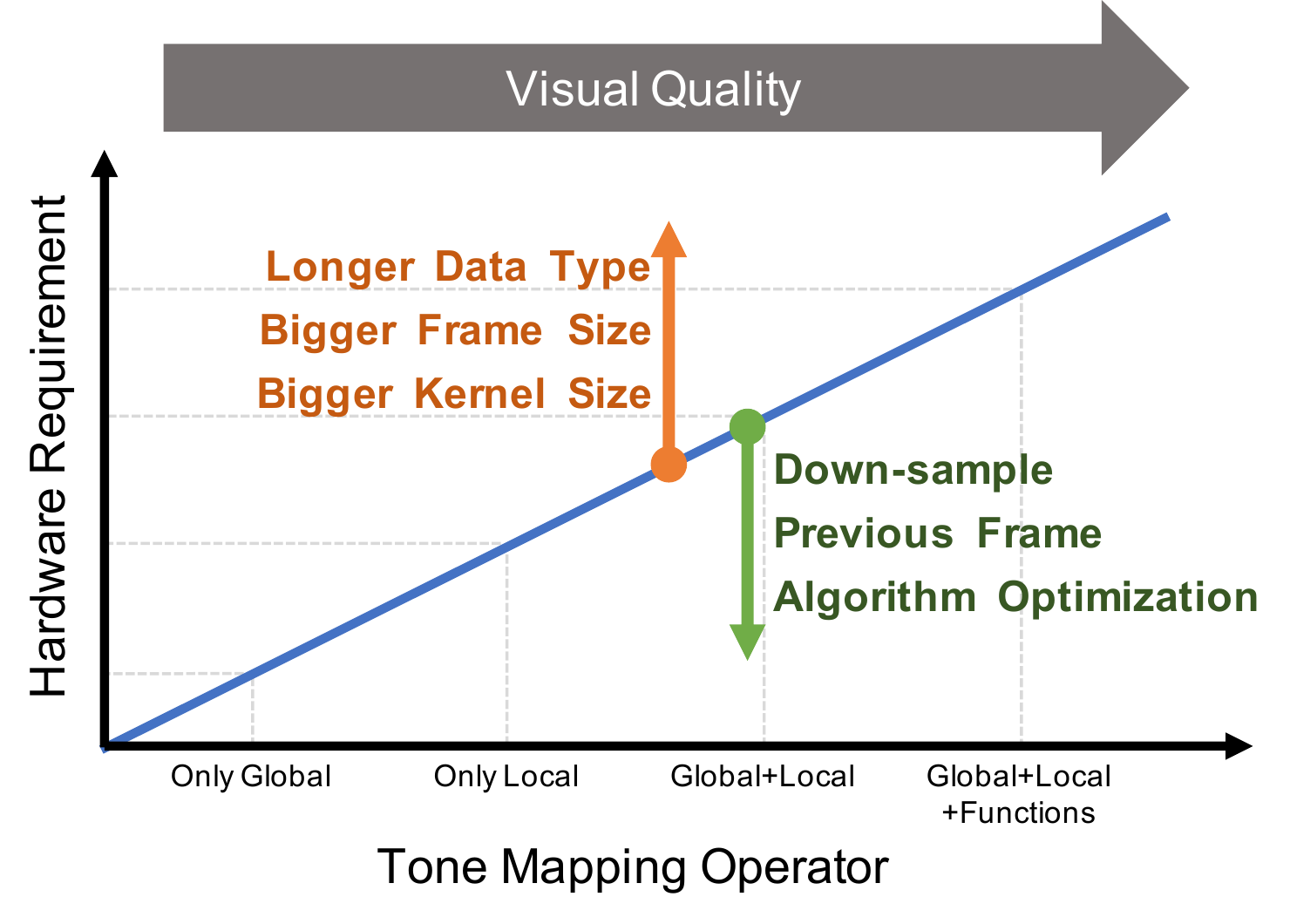}}
\caption{Trade-off: Visual quality versus hardware specification.}
\label{fig:HWRequire}
\end{figure}

Every image processing application seeks to achieve good output image quality, and TMOs are no different. Local TMO's are known to produce better images than global TMOs as they can reproduce both global and local contrast \cite{banterle2017advanced}. However, local TMOs are computationally more expensive than global TMOs and also may generate artifacts like halos around edges and amplify noise \cite{durand2002fast}. Therefore, additional functions are required along with local TMOs to reduce such artifacts. These local operations must be repeated over large amounts of data and usually demand substantial computational effort as shown in Fig .\ref{fig:HWRequire}, the visual quality does improve but at the expense of additional hardware. At the same time, depending on the data-type, image and kernel size will also increase the cost. As shown in Fig .\ref{fig:HWRequire}, cost can be reduced by applying techniques like operating the algorithm with previous frame/down-sampled images or employing other optimization techniques that can reduce computation or memory operations.        

As stated earlier, local contrast enhancement algorithms risk boosting noise. Many algorithm designers proposed various techniques to suppress noise. Eilertsen et al. presented a novel noise control with display-adaptivity to produce high contrast and detailed video given the display limitations \cite{eilertsen2015real}. Li et al. presented a new logarithmic CMOS sensor and a histogram-based tonemap operator which is derived from cumulative distribution function with an objective to suppress noise in the tone mapped image \cite{li2016novel}.

Ambalathankandy et al. in \cite{ambalathankandy2019fpga} implemented a halo reducing filter based on a Gaussian-like filter \cite{hore2014new}. In their TMO operator, halos were created around small bright features due to strong attenuation of neighboring pixels due to convolution operation with low-pass filtering. The hardware scheme for reducing such halos resulted in a very expensive implementation. 

Nosko et al. \cite{nosko2018color} proposed a ghost removal algorithm which significantly improves perceived quality of HDR image. This method is based on an earlier work of Grosch \cite{grosch2006fast}, and requires only simple arithmetic operations and thus it is suitable for implementation on FPGA. The ghost detection step is implemented before the HDR merging step. By constructing the ghostmap, marked pixel positions are treated differently from unmarked ones during the HDR merging.

Recently, Ambalathankandy et al. \cite{ambalathankandy2019adaptive} designed and implemented a LHE-based TMO that requires only one box filtering with a wide kernel. Their TMO algorithm uses two curves, one corresponding to the edge region and the other for gradation. As noises in gradation part are much more noticeable than in the edge and texture part. They use an alpha blending function to suppress noise in gradation region, and they include a halo control mechanism to manage light/dark halos individually using a simple weighting function on the bin-reduced histogram.

\section{Preferred Platform for Accelerating TMOs}

From Fig \ref{fig:paperNumber} we can observe that from early days FPGAs have been preferred platform for realizing real-time tone mapping applications. The main reasons are FPGAs popularity are:
 
\begin{itemize}
\item High Performance: The underlying FPGA fabric supports development of very deep pipeline architectures, with scope for wide parallel computational elements. This flexibility permits the designer to easily develop complex functionalities with strict timing constraints. Current generation of FPGAs can accelerate whole algorithm while processing full HD images or higher resolution images.
\item Re-programmability: This is one of the most attractive feature of FPGAs. A designer can iterate his design to make sure that he can tune his design to meet specific needs of targeted application without incurring additional cost.
\item Low Cost: The cost of developing a design on FPGA is comparatively cheaper, as many vendors provide customizable IP's and reference designs which can speed-up the development process.
\item Development Tools: Good support is available in the form of full development suite. Altera and Xilinx have streamlined development software for design and verification.
\item Flexibility for ASIC migration: FPGA proven design can be ported to structured ASICs which are available from many vendors, there by giving developers a faster route to market their products\cite{NEC, Fuji, eASIC}. 
\end{itemize}
\chapter{Future Perspectives and Conclusion}

Currently machine learning-based (ML) methods have become a very important tool to solve many computer vision tasks like image classification, face detection, and video analysis. As a future perspective we would like to leverage its potential by accelerating ML-based TMOs using hardware platforms. In this section we will explore the challenges and opportunities that we will encounter for such systems. Usually image processing tasks would require multiple convolution with fully connected layers, which are exorbitantly computationally intensive (for example, the operations in CNNs are over billion operations \cite{szegedy2015going}). Realizing such systems on resource constrained embedded systems would require very novel architectures and algorithms. FPGAs have been preferred platform for realizing CNN hardware accelerators for their following well-known features: re-programmability, low-power design features, and quick design time \cite{farabet2010hardware, zhang2015optimizing}. 

\begin{figure*}[!t]
\centerline{\includegraphics[width=\linewidth]{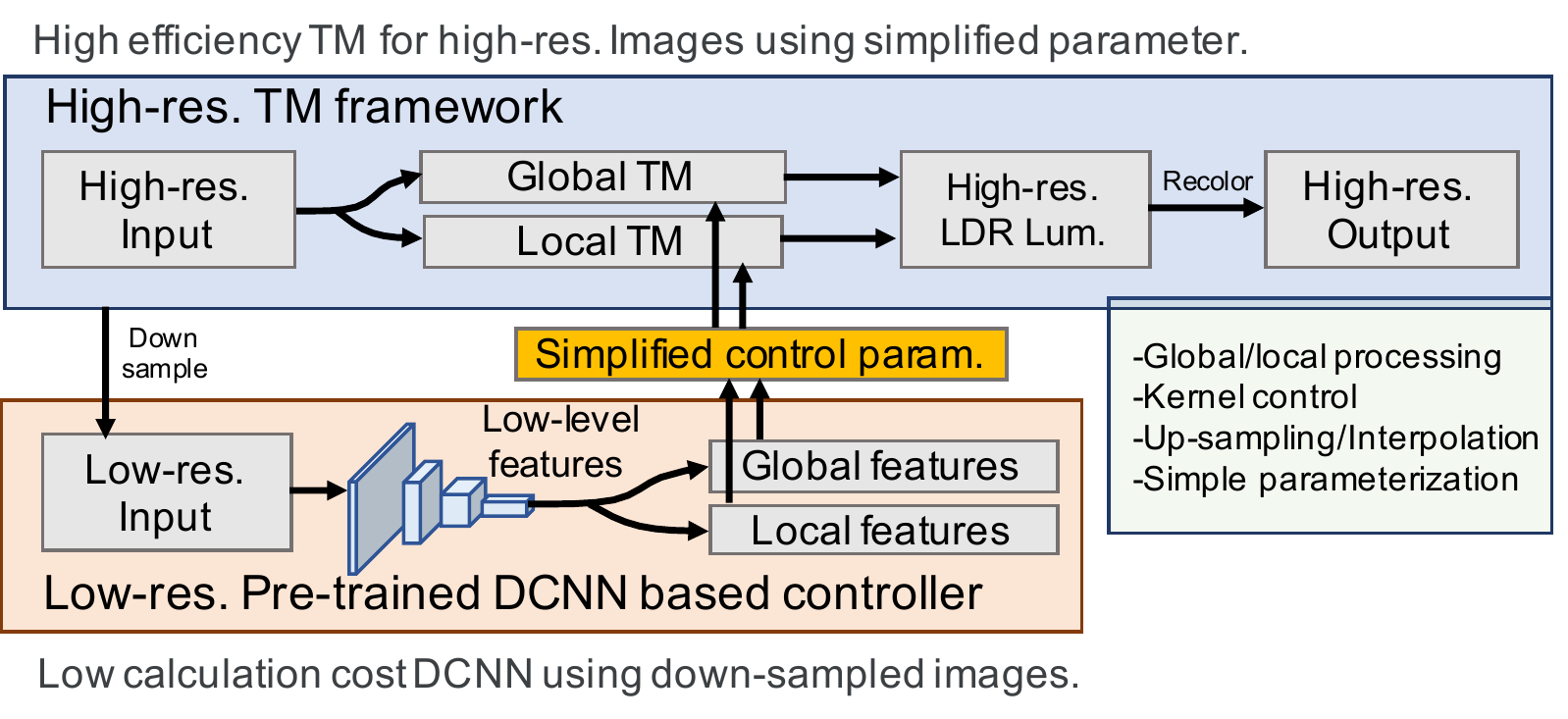}}
\caption{Future Perspective: Block diagram for a plausible machine learning-based TMO implementation on hardware.}
\label{fig:HDRnet}
\end{figure*}

Using Fig. \ref{fig:HDRnet} we demonstrate how Gharbi et al.’s HDR-Net like architectures are ideal baselines for realizing TMOs with DNN on hardware \cite{gharbi2017deep}. We find such designs are more hardware friendly because of the following features. First, the bilateral grid inspired architecture represents local tone-control as simple parametrized luminance grid in the space. Thus, the HDR-Net like high throughput design can approximate a tonemap system by using a  high-resolution guidance map which slices into the grid to produce a unique, interpolated, affine transform to be applied to each input pixel. Second, a lightweight DNN is vital for realizing hardware TMO system. Thus,  low-resolution DNN with down-sampling and optimum interpolation are important. Also, simple data transfer between DNN and TM system is required for reducing hardware load. In this scenario, output format of such DNN architectures becomes simple. Finally, good high resolution off-line dataset and training method through whole architecture is key for realizing this system.

In this survey we report a comprehensive list of about fifty tone mapping algorithms that have been implemented on hardware platforms like ASIC, FPGA and GPUs to accelerate the data intensive algorithms for real-time performance. Implementation of such algorithms are not usually uncomplicated, as hardware porting of their software equivalent may need to be redesigned for hardware-friendliness. This effort leads to various design challenges that are encountered during the hardware development. Usually the software algorithms are realized with floating-point data type and fixed-point conversion of the algorithms lead to loss of accuracy (image quality). In our literature survey we found that, various objective quality metrics have been used to demonstrate this distortion. For easy reference we have summarized all these objective metrics used in this survey. Finally, in this report we demonstrate the link between hardware cost and image quality thereby illustrating the underlying trade-off. This paper concludes with a discussion on the future perspectives of machine-learning based hardware TMO.

\bibliographystyle{plain}
\scriptsize
\bibliography{TMOSurvey}
\end{document}